\documentclass[a4paper,UKenglish,cleveref, autoref, thm-restate]{lipics-v2021}
\usepackage{marvosym}
\usepackage{todonotes}
\usepackage{numprint}
\npstyleenglish
\usepackage{placeins}
\usepackage{pgf}

\usepackage[ruled,vlined,linesnumbered,norelsize]{algorithm2e}
\DontPrintSemicolon

\SetKwSty{texttt}
\SetCommentSty{emph}

\title{Targeted Branching for the Maximum Independent Set Problem} 

\titlerunning{Targeted Branching for the MIS Problem} 

\author{Demian Hespe}{Karlsruhe Institute of Technology, Institute for
  Theoretical Informatics,
  Germany}{hespe@kit.edu}{https://orcid.org/0000-0001-6232-2951}{}

\author{Sebastian Lamm}{Karlsruhe Institute of Technology, Institute for
  Theoretical Informatics, Germany}{lamm@kit.edu}{https://orcid.org/0000-0001-7828-921X}{}

\author{Christian Schorr}{Karlsruhe Institute of Technology, Institute for
  Theoretical Informatics, Germany}{christian.schorr@student.kit.edu}{}{}

\authorrunning{D. Hespe and S. Lamm and C. Schorr} 

\Copyright{Demian Hespe and Sebastian Lamm and Christian Schorr} 



\category{} 

\relatedversion{} 

\acknowledgements{}

\nolinenumbers 

\hideLIPIcs  

\EventEditors{David Coudert and Emanuele Natale}
\EventNoEds{2}
\EventLongTitle{19th International Symposium on Experimental Algorithms (SEA 2021)}
\EventShortTitle{SEA 2021}
\EventAcronym{SEA}
\EventYear{2021}
\EventDate{June 7--9, 2021}
\EventLocation{Nice, France}
\EventLogo{}
\SeriesVolume{190}
\ArticleNo{17}

\usepackage{xspace}
\newcommand{\ie}{i.\,e.,\xspace}
\newcommand{\eg}{e.\,g.,\xspace}

\begin{document}

\maketitle

\begin{abstract}
  Finding a maximum independent set is a fundamental NP-hard problem that is used in many real-world applications.
Given an unweighted graph, this problem asks for a maximum cardinality set of pairwise non-adjacent vertices.
In recent years, some of the most successful algorithms for solving this problem are based on the branch-and-bound or branch-and-reduce paradigms.
In particular, branch-and-reduce algorithms, which combine branch-and-bound with reduction rules, have been able to achieve substantial results, solving many previously infeasible real-world instances.
These results were to a large part achieved by developing new, more practical reduction rules.
However, other components that have been shown to have a significant impact on the performance of these algorithms have not received as much attention.
One of these is the branching strategy, which determines what vertex is included or excluded in a potential solution.
Even now, the most commonly used strategy selects vertices solely based on their degree and does not take into account other factors that contribute to the performance of the algorithm.

In this work, we develop and evaluate several novel branching strategies for both branch-and-bound and branch-and-reduce algorithms.
Our strategies are based on one of two approaches which are motivated by existing research. 
They either (1) aim to decompose the graph into two or more connected components which can then be solved independently, or (2) try to remove vertices that hinder the application of a reduction rule which can lead to smaller graphs.
Our experimental evaluation on a large set of real-world instances indicates
that our strategies are able to improve the performance of the state-of-the-art
branch-and-reduce algorithm by Akiba and Iwata. To be more specific,
our reduction-based packing branching rule is able to outperform the default
branching strategy of selecting a vertex of highest degree on $65\%$ of all
instances tested.
Furthermore, our decomposition-based strategy based on edge cuts is able to achieve a speedup of $2.29$ on sparse networks ($1.22$ on all instances).

\end{abstract}

\section{Introduction}

An \emph{independent set} of a graph $G = (V,E)$ is a set of vertices $I \subseteq V$ of $G$ such that no two vertices in this set are adjacent.
The problem of finding such an independent set of maximum cardinality, the \emph{maximum independent set problem}, is a fundamental NP-hard problem~\cite{Garey1974}.
Its applications cover a wide variety of fields including computer graphics \cite{CG}, network analysis \cite{NW}, route planning \cite{RP} and computational biology \cite{BIO1, BIO2}.
In computer graphics for instance, large independent sets can be used to optimize the traversal of mesh edges in a triangle mesh.
Further applications stem from its complementary problems minimum vertex cover and maximum clique.

One of the best known techniques for finding maximum independent sets, both in
theory~\cite{XiaoNagamochi, ChenXiaKanj} and practice \cite{AkibaIwata}, are
\emph{data reduction algorithms}.
These algorithms apply a set of reduction rules to decrease the size of an instance while maintaining the ability to compute an optimal solution afterwards.
A recently successful type of data reduction algorithm is so-called
\emph{branch-and-reduce algorithms}~\cite{AkibaIwata,WGYC}, which exhaustively
apply a set of reduction rules to compute an irreducible graph.
If no further rule can be applied, the algorithm branches into (at least) two
smaller subproblems, which are then solved recursively.
To make them more efficient in practice, these algorithms also make use of problem-specific upper and lower bounds to quickly prune the search space.

Due to the practical impact of data reduction, most of the research aimed at
improving the performance of branch-and-reduce algorithms so far has been
focused on either proposing more practically efficient special cases of already
existing rules~\cite{ChangKern,dahlum2016accelerating}, or maintaining
dependencies between reduction rules to reduce unnecessary
checks~\cite{alsahafy2020computing,hespe2019scalable}.
However, improving other aspects of branch-and-reduce has been shown to benefit its performance~\cite{plachetta2021sat}.
The branching strategy in particular has been shown to have a significant impact on the running time~\cite{AkibaIwata}.
Up to now, the most frequently used branching strategy employed in many state-of-the-art solvers selects branching vertices solely based on their degree.
Other factors, such as the actual reduction rules used during the algorithm are rarely taken into account.
However, recently there have been some attempts to incorporate such branching
strategies for other problems such as finding a maximum $k$-plex~\cite{gao2018exact}.

\subsection{Contribution}
In this paper, we propose and examine several novel strategies for selecting branching vertices.
These strategies follow two main approaches that are motivated by existing research: (1) Branching on vertices that decompose the graph into several connected components that can be solved independently.
Solving components individually has been shown to significantly improve the performance of branch-and-reduce in practice, especially when the size of the largest component is small~\cite{alsahafy2020computing}.
(2) Branching on vertices whose removal leads to reduction rules becoming applicable again.
In turn, this leads to a smaller reduced graph and thus improved performance.
For each approach we present several concrete strategies that vary in their complexity.
Finally, we evaluate their performance by comparing them to the aforementioned default strategy used in the state-of-the-art solver by Akiba and Iwata~\cite{AkibaIwata}.
For this purpose we make use of a wide spectrum of instances from different graph classes and applications.
Our experiments indicate that our strategies are able to find an optimal
solution faster than the default strategy on a large set of instances.
In particular, our reduction-based packing rule is able to outperform the default strategy on $65\%$ of all instances.
Furthermore, our decomposition-based strategies achieve a speedup of $1.22$ (over the default strategy) over all instances.
A more detailed explanation of a previous version of this work
can be found in Schorr's Bachelor's thesis~\cite{schorr2020improved}.

\section{Preliminaries}
Let $G=(V,E)$ be an undirected graph, where $V = \{0, \ldots, n-1\}$ is a set of $n$ vertices and $E \subseteq  \{\{u,v\} \mid u,v \in V\}$ is a set of $m$ edges. 
We assume that $G$ is \emph{simple}, \ie it has no self loops or multi-edges.
The \emph{(open) neighborhood} of a vertex $v \in V$ is denoted by $N(v) = \{u \mid \{v,u\} \in E\}$.
Furthermore, we denote the \emph{closed neighborhood} of a vertex by $N[v]=N(v) \cup \{v\}$.
We define the open and closed neighborhood of a set of vertices $U \subseteq V$
as $N(U) = \cup_{u \in U} N(u) \setminus U$ and $N[U] = N(U) \cup U$, respectively.
The \emph{degree} of a vertex $v \in V$ is the size of its neighborhood $d(v) =
|N(v)|$ and $\Delta =
\max_{v \in V} \{d(v)\}$.
For a vertex $v \in V$, we further define $N^2(v) = N(N(v))$.

For a subset of vertices $V_S \subseteq V$, the \emph{(vertex-)induced subgraph}
$G[V_S] = (V_S, E_S)$ is given by restricting the edges of $G$ to vertices of $V_S$, \ie $E_S = \{\{u,v\} \in E \mid u,v \in V_S\}$.
Likewise, for a subset of edges $E_S \subseteq E$, the \emph{edge-induced
  subgraph} $G[E_S] = (V_S, E_S)$ is given by restricting the vertices of $G$ to the endpoints of edges in $E_S$, \ie $V_S = \{u,v \in V \mid \{u,v\} \in E_S\}$.
  For a subset of vertices $U \subset V$, we further define $G - U$ as the induced subgraph $G[V \setminus U]$.

A \emph{path} $P=(v_1, \ldots, v_k)$ of length $k$ is a sequence of $k$ distinct vertices in $G$ such that $\{v_i, v_{i+1}\} \in E$ for all $i \in \{1, \ldots, k-1\}$.
A subgraph of $G$ induced by a maximal subset of vertices that are connected by a path is called a \emph{connected component}.
Furthermore, a graph that only contains one connected component is called \emph{connected}.
Likewise, a graph with more than one connected component is called \emph{disconnected}.
A subset $S \subset V$ of a connected graph $G$ is called a \emph{vertex separator} if the removal of $S$ from $G$ makes the graph disconnected.

An \emph{independent set} of a graph is a subset of vertices $I \subseteq V$ such that no two vertices of $I$ are adjacent. 
A \emph{maximum independent set} (MIS) is an independent set of maximum cardinality.
Closely related to independent set are vertex covers and cliques.
A \emph{vertex cover} is a set of vertices $C \subseteq V$ such that for each edge $\{u,v\} \in E$ either $u$ or $v$ is contained in $C$.
The complement of a (maximum) independent set of a graph $G$ is a \emph{(minimum) vertex cover} (MVC) of $G$.
A \emph{clique} is a subset of vertices $K \subseteq V$ such that all vertices of $K$ are adjacent to each other, \ie $\forall u,v \in K: \{u,v\} \in E$.
Finally, a (maximum) independent set of a graph $G$ is a \emph{(maximum) clique} (MC) in the complement graph $\overline{G} = (V, \overline{E})$, where $\overline{E} = \{\{u,v\} \mid \{u,v\} \not\in E\}$.

\section{Related Work}

The most commonly used branching strategy for MIS and MVC is to select a vertex
of maximum degree. Fomin et al.~\cite{Fomin} show that using a vertex of maximum
degree that also minimizes the number of edges between its neighbors is optimal
with respect to their complexity measure. The algorithm by Akiba and Iwata~\cite{AkibaIwata}
(which we augment with our new branching rules) also uses this strategy. Akiba and Iwata also
compare this strategy to branching on a vertex of minimum degree and a random
vertex. They show that both of these perform significantly worse than branching
on a maximum degree vertex.

Xiao and Nagamochi~\cite{XiaoNagamochi} also use
this strategy in most cases. For dense subgraphs, however, they use an edge
branching strategy: They branch on an edge $\{u, v\}$ where $|N(u) \cap N(v)|$
is sufficiently large (depending on the maximum degree of the graph) by
excluding both $u$ and $v$ in one branch and applying the alternative reduction
(see Section~\ref{sec:almost_funnels}) to $\{u\}$ and $\{v\}$ in the other branch.

Bourgeois et al.~\cite{Bourgeois} use maximum degree branching as long as there are vertices of degree at least five. Otherwise, they utilize specialized algorithms to solve subinstances with an average degree of three or four. Those algorithms perform a rather complex case analysis to find a suitable branching vertex. The analysis is based on exploiting structures that contain 3- or 4-cycles. Branching on specific vertices in such structures often enables further reduction rules to be applied.

Chen et al.~\cite{ChenXiaKanj} use a notion of \emph{good pairs} that are advantageous
for branching. They chose these good pairs by a set of rules which are omitted
here. They combine these with so-called \emph{tuples} of a set of vertices and the
number of vertices from this set that have to be included in an MIS. This
information can be used when branching on a vertex
contained in that set to remove further vertices from the graph. Akiba and
Iwata~\cite{AkibaIwata} use the same concept in their \emph{packing} rule. Chen
et al. combine good pairs, tuples and high degree vertices for their branching strategy.

Most algorithms for MC (e.g. \cite{Color,DBLP:journals/ieicet/TomitaSHW13}) compute a
greedy coloring and branch on vertices with a high coloring number.
More sophisticated MC algorithms use MaxSAT encodings to prune the set of
branching vertices~\cite{LiFangXu,LiJiang,LiQuan}. Li et al.~\cite{LiMaxSat}
combine greedy coloring and MaxSAT reasoning the further reduce the number of
branching vertices.

Another approach used for MC is using the \emph{degeneracy order} $v_1 < v_2 <
\dots < v_n$ where $v_i$ is a vertex of smallest degree in $G - \{v_1, \dots
v_{i-1}\}$. Carraghan and Pardalos~\cite{CarraghanPardalos} present an algorithm
that branches in descending degeneracy order. Li et al.~\cite{LiFangXu}
introduce another vertex ordering using iterative maximum independent set
computations (which might be easier than MC on some graphs) and breaking ties
according to the degeneracy order.

The algorithm by Akiba and Iwata~\cite{AkibaIwata} is a so-called
\emph{branch-and-reduce} algorithm: It repeatedly reduces the instance size by a
set of polynomial-time reduction rules and then branches on a vertex once no
more reduction rules can be applied. Since branching removes at least one vertex
from the graph, more reduction rules might be applicable afterwards. The
set of reductions used in their algorithm is relatively large and not covered
completely here. However, some reduction rules are explained in
Section~\ref{sec:reduction_branching} where we show how to target particular reduction
rules when branching. Akiba and Iwata apply the reduction rules in a predefined
order. For each rule, their algorithm iterates over all
vertices in the graph and checks whether the rule
can be applied. If a rule is applied successfully, this process is
restarted from the first reduction rule. In order to prune the search space, bounds on the largest
possible independent set of a branch are computed. They implement three different
methods for determining upper bounds: clique cover, LP relaxation and cycle
cover. Additionally, they employ special reduction rules that can be applied
during branching. Another optimization done by their algorithm is to solve
connected components separately. We utilize this in
Section~\ref{sec:decomposition_branching} where we introduce branching rules
that decompose the graph into connected components. We use this algorithm
as the base implementation to test our new branching strategies.

\section{Decomposition Branching}\label{sec:decomposition_branching}
Our first approach to improve the default branching strategy found in many state-of-the-art algorithms (including that of Akiba and Iwata~\cite{AkibaIwata}) is to decompose the graph into several connected components.
Subsequently, processing these components individually has been shown to improve the performance of branch-and-reduce in practice~\cite{alsahafy2020computing}.
To this end, we now present three concrete strategies with varying computational
complexity: articulation points, edge cuts and nested dissections.

\subsection{Articulation Points}
First, we are concerned with finding single vertices that are able to decompose a graph into at least two separated components.
Such points are called \emph{articulation points} (or cut vertices).
Articulation points can be computed in linear time $\mathcal{O}(n+m)$ using a simple depth-first search (DFS) algorithm (see Hopcroft and Tarjan~\cite{hopcroft1973algorithm} for a detailed description).
In particular, a vertex $v$ is an articulation point if it is either the root of
the DFS tree and has at least two children or any non-root vertex that has a child $u$, such that no vertex in
the subtree rooted at $u$ has a back edge to one of the ancestors of
$v$. 

\begin{figure}[t]
  \centering
  \includegraphics[scale=1]{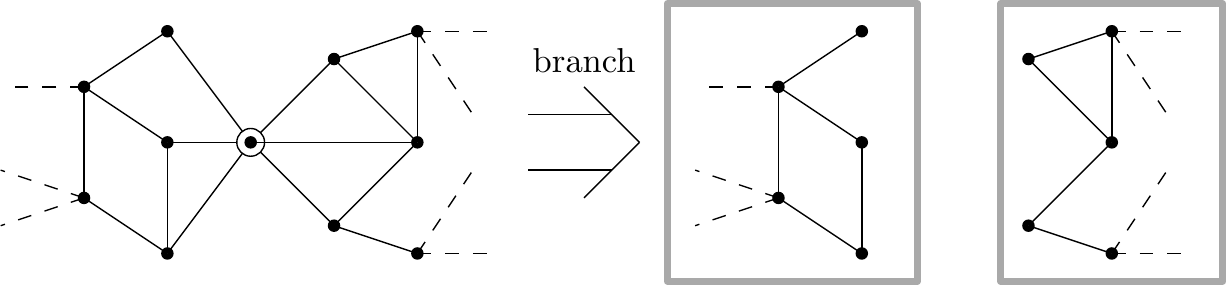}
  \caption{Branching on an articulation point (circled vertex)
    decomposes the graph into two connected components (gray boxes) that can be
    solved independently. The graphic shows the branch in which the vertex is
    excluded from the independent set.}\label{fig:articulation_points}
\end{figure}

For our first branching strategy we maintain a set of articulation points $A \subseteq V$.
When selecting a branching vertex, we first discard all invalid vertices from
$A$, \ie vertices that were removed from the graph by a preceding data reduction step.
If this results in $A$ becoming empty, a new set of articulation points is computed on the current graph in linear time.
However, if no articulation points exist, we select a vertex based on the default branching strategy.
Otherwise, if $A$ contains at least one vertex, an arbitrary one from $A$ is selected as the branching
vertex. Figure~\ref{fig:articulation_points} illustrates branching on an
articulation point.

Even though this strategy introduces only a small (linear) overhead, finding articulation points can be rare depending on the type of graph.
This results in the default branching strategy being selected rather
frequently.
Furthermore, our preliminary experiments indicate that articulation points are rarely found at higher depth.
However, due to their low overhead, we can justify searching for them whenever
$A$ becomes empty.

\subsection{Edge Cuts}
To alleviate the restrictive nature of finding articulation points, we now propose a more flexible branching strategy based on \emph{(minimal) edge cuts}.
In general, we aim to find small vertex separators, \ie a set of vertices whose removal disconnects the graph.
We do so by making use of minimum edge cuts.

A \emph{cut} $(S,T)$ is a partitioning of $V$ into two sets $S$ and $T=V\setminus S$.
Furthermore, a cut is called minimum if its \emph{cut set} $C = \{\{u,v\} \in E \mid u \in S, v \in T\}$ has minimal cardinality.
However, in practice, finding minimum cuts often yields trivial cuts with either $S$ or $T$ only consisting of a single vertex with minimum degree. 
Thus, we are interested in finding \emph{$s$-$t$-cuts}, \ie cuts where $S$ and $T$ contain specific vertices $s,t \in V$.
Finding these cuts can be done efficiently in practice, \eg using a preflow push algorithm~\cite{goldberg1988new}.
However, selecting the vertices $s$ and $t$ to ensure reasonably balanced cuts can be tricky.
Natural choices include random vertices, as well as vertices that are far apart in terms of their shortest path distance.
However, our preliminary experiments indicate that selecting random vertices of maximum degree
 for $s$ and $t$ seems to produce the best results.
Finally, to derive a vertex separator from a cut, one can compute an MVC on the bipartite graph induced by the cut set, \eg using the Hopcroft-Karp algorithm~\cite{hopcroft1973n}.
This separator can then be used to select branching vertices. In particular,
we continuously branch on vertices from the separator.

Overall, our second strategy works similar to the first one: We maintain a set of possible branching vertices that were selected by computing a minimum $s$-$t$-cut and turning it into a vertex separator.
Vertices that were removed by data reduction are discarded from this set and once it is empty a new cut computation is started.
However, in contrast to the first strategy, finding a set of suitable branching vertices is much more likely.
In order to avoid separators that contain too many vertices, and thus would require too many branching steps to disconnect the graph, we only keep those that do not exceed a certain size and balance threshold.
The specific values for these threshold are presented in Section~\ref{sec:algo_conf}.
Finally, if no suitable separator is found, we use the default branching strategy.
Furthermore, in this case we do not try to find a new separator for a fixed number of branching steps as finding one is both unlikely and costly.

\subsection{Nested Dissection}
Both of our previous strategies dynamically maintain a set of branching vertices.
Even though this comes at the advantage that most of the computed vertices remain viable candidates for some branching steps, it introduces a noticeable overhead.
To alleviate this, our last strategy uses a static ordering of possible branching vertices that is computed once at the beginning of the algorithm.
For this purpose we make use of a \emph{nested dissection ordering}~\cite{george1973nested}.

A nested dissection ordering of the vertices of a graph $G$ is obtained by recursively computing balanced bipartitions $(A,B)$ and a vertex separator $S$, that separates $A$ and $B$.
The actual ordering is then given by concatenating the orderings of $A$ and $B$ followed by the vertices of $S$.
Thus, if we select branching vertices based on the reverse of a nested dissection
ordering, we continuously branch on vertices that disconnect the graph into balanced partitions. 
We compute such an ordering once, after finishing the initial data reduction phase.

There are two main optimizations that we use when considering the nested dissection ordering.
First, we limit the number of recursive calls during the nested dissection
computation, because we noticed that vertices at the end of the ordering seldom lead to a decomposition of the graph. 
This is due to the graph structure being changed by data reduction which can lead to separators becoming invalid.
Furthermore, similar to the edge-cut-based strategy, we limit the size of
separators considered during branching using a threshold.
Again, this is done to ensure that we do not require too many branching steps to decompose the graph.
The specific value for this size threshold is given in Section~\ref{sec:algo_conf}.
If any separator in the nested dissection exceeds this threshold, we use the default branching strategy.

\section{Reduction Branching}\label{sec:reduction_branching}
Our second approach to selecting good branching vertices is to choose a vertex
whose removal will enable the application of new reduction rules. During every
reduction step we find a list of candidate vertices to branch on. The following
sections will demonstrate how we identify such branching candidate vertices with little
computational overhead in practice. To be self contained we will also repeat the reduction
rules used here but omit any proofs that can be found in the original
publications. Out of the candidates found we then select a vertex of
maximum degree. If the degree of all candidate vertices lies below a threshold
(defined in Section~\ref{sec:algo_conf}) or no candidate vertices were found, we fall back to branching on a vertex
of maximum degree. The rational here is that a vertex of large degree changes
the structure of the graph more than a vertex of small degree even if that
vertex is guaranteed to enable the application of a reduction rule. Also, our
current strategies (except the packing-based rule in Section~\ref{sec:almost_packing}) only enable the application of the targeted reduction rule in
the branch that excludes the vertex from the independent set, the
\emph{excluding branch}. However, in the
case that includes it into the independent set (\emph{including branch}) all neighbors are removed from
the graph as well because they already have an adjacent vertex in the solution.
Thus, in both branches multiple vertices are removed.

We also performed some preliminary experiments with storing the candidate vertices in
a priority queue without resetting after every branch.
However, changes were too frequent for this approach to be faster because of
the high amount of priority queue operations.

\subsection{Almost Twins}

The first reduction we target is the \emph{twin} reduction by Xiao and Nagamochi~\cite{XiaoUnconfined}:

\begin{definition}(Twins~\cite{XiaoUnconfined})
  In a graph $G=(V,E)$ two vertices $u$ and $v$ are called twins if $N(u) = N(v)$ and $d(u) = d(v) = 3$.
\end{definition}

\begin{theorem} (Twin Reduction~\cite{XiaoUnconfined}) In a graph $G=(V,E)$ let
  vertices $u$ and $v$ be twins. If there is an edge among $N(u)$, then there is
  always an MIS that includes $\{u,v\}$ and therefore
  excludes $N(u)$. Otherwise, let $G'=(V',E')$ be the graph with $V'=(V\setminus
  N[\{u,v\}])\cup\{w\}$ where $w\notin V$ and $E'=(E\cap\binom{V'}{2})\cup
  \{\{w,x\}\;|\;x\in N^2(u)\}\}$ and let $I'$ be an MIS in $G'$. Then, 
  $
  I=\begin{cases}
    I'\cup \{u,v\} & \text{, if }w\notin I'\\
    (I'\setminus \{w\})\cup N(u) & \text{, else} 
  \end{cases}
  $
  is an MIS in $G$.
\end{theorem}

We now define \emph{almost twins} as follows:
\begin{definition} (Almost Twins)
  In a graph $G=(V,E)$ two non adjacent vertices $u$ and $v$ are called almost twins if $d(u) = 4$, $d(v) = 3$ and $N(v)\subseteq N(u)$ (\ie $N(u) = N(v) \cup \{w\}$). 
\end{definition}

\begin{figure}[t]
  \centering
  \includegraphics[scale=1]{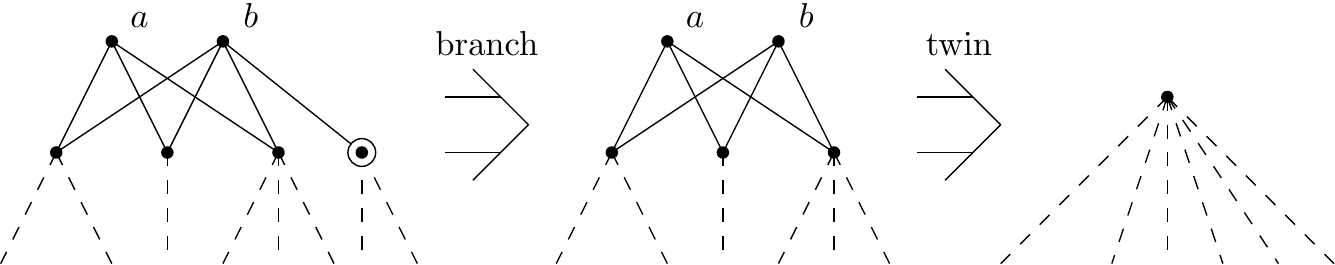}
  \caption{Vertices $a$ and $b$ are almost twins. After branching on
  the circled vertex they become twins (in the excluding branch) and can
  be reduced.}\label{fig:twin}
\end{figure}

Clearly, after removing $w$, $u$ and $v$ are twins so we can apply the twin
reduction. Finding almost twins can be done while searching for twins: The
original algorithm checks for each vertex $v$ of degree-$3$ whether there is a
vertex $u \in N^2(v)$ with $d(u) = 3$ and $N(u) = N(v)$. We augment this
algorithm by simultaneously also searching for $u \in N^2(v)$ with $d(u) = 4$
and $N(v) \subset N(u)$. This induces about the same computational cost for
degree-$4$ vertices in $N^2(v)$ as for degree $3$ vertices. While there might be
instances where this causes high overhead, we expect the practical slowdown to
be small. Figure~\ref{fig:twin} illustrates branching for almost twins. 

\subsection{Almost Funnels}\label{sec:almost_funnels}

Next, we consider the \emph{funnel} reduction which is a special case of the
\emph{alternative} reduction by Xiao and Nagamochi~\cite{XiaoUnconfined}:

\begin{definition} (Alternative Sets~\cite{XiaoUnconfined})
	In a graph $G=(V,E)$ two non empty, disjoint subsets $A,B\subseteq V$ are
    called alternatives if $|A| = |B|$ and there is an MIS $I$ in $G$ such that $I\cap(A\cup B)$ is either $A$ or $B$.
\end{definition}

\begin{theorem} (Alternative Reduction~\cite{XiaoUnconfined})
	In a graph $G=(V,E)$ let $A$ and $B$ be alternative sets. Let $G'=(V', E')$
    the graph with $V' = V\setminus(A\cup B\cup (N(A)\cap N(B)))$ and $E' =
    \{\{u,v\} \in E \mid u,v \in V'\} \cup \{ \{u,v\}\;|\; u\in
    N(A)\setminus N[B], v\in N(B)\setminus N[A] \}$ and let $I'$ be an MIS in $G'$. Then,
	$
    I=\begin{cases}
      I'\cup A & \text{, if } (N(A)\setminus N[B]) \cap I' = \emptyset
      \\
      I'\cup B & \text{, else}
	\end{cases}
    $
	is an MIS in $G$.
\end{theorem}

Note that the alternative reduction adds new edges between existing
vertices of the graph which might not be beneficial in every case. To counteract this, the algorithm
by Akiba and Iwata~\cite{AkibaIwata} only uses special cases, one of which is
the funnel reduction:

\begin{definition} (Funnel~\cite{XiaoUnconfined})
	In a graph $G=(V,E)$ two adjacent vertices $u$ and $v$ are called funnels if $G_{N(v)\setminus\{u\}}$ is a complete graph, i.e, if $N(v)\setminus\{u\}$ is a clique.
\end{definition}
\begin{theorem} (Funnel Reduction~\cite{XiaoUnconfined}) In a graph $G=(V,E)$ let $u$ and $v$ be funnels. Then, $\{u\}$ and $\{v\}$ are alternative sets.	
\end{theorem}

Again, we define a structure that is covered by the funnel reduction after
removal of a single vertex:

\begin{definition}(Almost Funnel)
  In a graph $G=(V,E)$ two adjacent vertices $u$ and $v$ are called almost funnels if $u$ and $v$ are not funnels and there is a vertex $w$ such that $N(v)\setminus\{u,w\}$ induces a clique.
\end{definition}

By removing $w$, $u$ and $v$ become funnels. The original funnel algorithm
checks whether $u$ and $v$ are funnels by iterating over the vertices in $N(v) \setminus \{u\}$
and checking whether they are adjacent to \emph{all} previous vertices. Once a vertex
is found that is not adjacent to all previous vertices, the algorithm concludes
that $u$ and $v$ are not funnels and terminates. We augment this algorithm by
not immediately terminating in this case. Instead, we consider the following two
cases: Either the current vertex $w$ is not adjacent to at least two of the
previous vertices. In this case, we can check whether $N(v) \setminus \{u, w\}$
induces a clique. In the second case, $w$ is adjacent to all but one previous
vertex $w'$. In this case, both $w$ and $w'$ might be candidate branching
vertices. Thus, we check whether $N(v) \setminus \{u, w\}$ or $N(v) \setminus
\{u, w'\}$ induce a clique. This adds up to two additional clique checks (of
slightly smaller size) to the one clique check in the original algorithm.

\subsection{Almost Unconfined}\label{sec:almost_unconfined}

The core idea of the \emph{unconfined} reduction by Xiao and
Nagamochi~\cite{XiaoUnconfined} is to detect vertices not required for an MIS
that can therefore be removed from the graph by algorithmically
contradicting the assumption that every MIS contains the vertex.

\begin{definition} (Child, Parent~\cite{XiaoUnconfined}) In a graph $G=(V,E)$ with an \textit{independent set} $I$, a vertex $v$ is called a child of $I$ if $|N(v)\cap I| = 1$ and the unique neighbor of $v$ in $I$ is called the parent of $v$.
\end{definition}

Algorithm~\ref{alg:unconf} shows the algorithm used by Akiba and
Iwata~\cite{AkibaIwata} to detect so called \textit{unconfined}
vertices.

\begin{algorithm}[t]
	\caption{Unconfined -- Xiao and Nagamochi
      \cite{XiaoUnconfined}}\label{alg:unconf}
	\SetKwFunction{U}{Unconfined}
	\DontPrintSemicolon
	
	\KwIn{A graph $G$, a vertex $v$}
	\U{G, v}	
	\Begin{
		$S \leftarrow \{v\}$\;
		\While{$S\text{ has child }u\text{ with } |N(u)\setminus N[S]|\leq 1$}{
			\eIf{$|N(u)\setminus N[S]| = 0$}{\Return true}{
				$\{w\}\leftarrow N(u)\setminus N[v]$\tcp*{by assumption $w$ also has to}
				$S\leftarrow S\cup\{w\}$\tcp*{be contained in every MIS}
			}
		}
		
		\Return $\text{false}$
		
	}
	
	\KwOut{true if $v$ is unconfined, false otherwise}
	
\end{algorithm}

\begin{theorem}(Unconfined Reduction~\cite{XiaoUnconfined}) In a graph $G=(V,E)$,
  if Algorithm \ref{alg:unconf} returns true for an unconfined vertex $v$, then
  there is always an MIS that does not contain~$v$.
\end{theorem}

Again, we define a vertex to be almost unconfined:

\begin{definition} (Almost Unconfined)
  In a graph $G=(V,E)$ a vertex $v$ is called almost unconfined if $v$ is not unconfined but there is a vertex $w$ such that $v$ is unconfined in $G-\{w\}$.
\end{definition}

Here, we only present an augmentation that detects \emph{some} almost
unconfined vertices. In particular, if at any point during the algorithm there
is only \emph{one} extending child, i.e. a child $u$ of $S$ with $N(u)\setminus
N[S] = \{w\}$, then removal of $w$ makes $v$ unconfined. During
Algorithm~\ref{alg:unconf} we collect all these vertices $w$ and add them to the
set of candidate branching vertices if the algorithm cannot already remove $v$.
This only adds the overhead of temporarily storing the potential candidates and
adding them to the actual candidate list if $v$ is not removed.


\subsection{Almost Packing}\label{sec:almost_packing}

The core idea behind the packing rule by Akiba and Iwata \cite{AkibaIwata} is
that when the exluding branch of a vertex $v$ is selected, one
can assume that no maximum independent set contains $v$. Otherwise, if there is
a maximum independent set that contains $v$, the algorithm finds it in the including branch of $v$. Based on the assumption that no maximum independent
set includes a vertex $v$, constraints for the remaining vertices can be
derived. For example, a maximum independent set that does not contain $v$ has to
include at least two neighbors of $v$. The corresponding constraint is
$\sum_{u\in N(v)}x_u \geq2$, where $x_u$ is a binary variable that indicates
whether a vertex is included in the current solution. Otherwise, we will find a
solution of the same size in the branch including $v$. The algorithm creates such
constraints when branching or reducing, and updates them accordingly during the data reductions and branching steps. When a vertex $v$ is eliminated from the graph, $x_v$ gets removed from all constraints. If $v$ is included into the current solution, the corresponding right sides are also decreased by one.

A constraint $\sum_{v\in S\subset V}x_v\geq k$ can be utilized in two
reductions. Firstly, if $k$ is equal to the number of variables $|S|$, all
vertices from $S$ have to be included into the current solution. If there are
edges between vertices from $S$, then no valid solution can include all vertices
from $S$, so the branch is pruned. Secondly, if there is a vertex $v$ such that $|S|-|N(v)\cap S| < k$, then $v$ has to be excluded from the current solution. If $k > |S|$, the constraint can not be fulfilled and the current branch is pruned.

In our branching strategy we target both reductions. If there is a constraint $\sum_{v\in S\subset V}x_v\geq k$, where $|S| = k + 1$, excluding any vertex of $S$ from the solution or including a vertex of $S$ that has one neighbor in $S$ enables the first reduction. Thus, we consider all vertices in $S$ for branching. Note that including a vertex from $S$ that has more than one neighbor in $S$ makes the constraint unfulfillable and the branch is pruned.

If there is a constraint $\sum_{v\in S\subset V}x_v\geq k$ and a vertex $v$, such that $k = |S|-|N(v)\cap S|$, excluding any vertex of $S\setminus N(v)$ from the solution or including a vertex of $S\setminus N(v)$ that has at least one neighbor in $S\setminus N(v)$ enables the second reduction. Thus, we consider all vertices in $S\setminus N(v)$ for branching.

Note that in contrast to our previous reduction-based branching rules, packing
reductions can also be applied in the including branch in many cases.

Detecting these branching candidates can be done with small constant overhead whilst performing the packing reduction.

\section{Experimental Evaluation}

In this section, we present the results of our experimental evaluation. Tables
and figures here show aggregated results. For
detailed results for all of our algorithms across all instances, see Appendix~\ref{app:detailed_results}.

\subsection{Experimental Environment}
We augment a C++-adaptation of the algorithm by Akiba and
Iwata~\cite{AkibaIwata} with our branching strategies and compile it
with g++ 9.3.0 using full optimizations (\texttt{-O3}). Our
code is publicly available on GitHub\footnote{\url{https://github.com/Hespian/CutBranching}}. We execute all our experiments on a machine with 4 8-core Intel Xeon E5-4640 CPUs
(2.4~GHz) and 512 GiB DDR3-PC1600 RAM running Ubuntu 20.04.1 with Linux Kernel 5.4.0-64. To speed up our experiments we use two
identical machines and run at most 8 instances at once on the
same machine (using the same machine for all algorithms on a specific instance).
All numbers reported are arithmetic means of three runs with a timeout of ten
hours.

\subsection{Algorithm Configuration}\label{sec:algo_conf}
We use a C++ adaptation of the implementation by Akiba and
Iwata~\cite{AkibaIwata} in its default configuration as a basis for our algorithm. During preliminary experiments we found
suitable values for the parameters of
our techniques. These experiments were run on a subset of our total instance
set. We use the geometric mean over all instances of the speedup over the
default branching strategy as a basis for the following decisions: for the technique based on
edge cuts, we only use cuts that contain at most 25 vertices and where the smaller side of
the cut contains at least ten percent of the remaining vertices. If no suitable separator is found, we skip ten branching steps. For computing nested dissections, we use 
InertialFlowCutter~\cite{gottesburen2019faster}
with the KaFFPa~\cite{DBLP:conf/wea/SandersS13}
backend. The KaFFPa partitioner is configured to use the \emph{strong} preset
with a fixed seed of $42$. For branching, we use
three levels of nested dissections with a minimum balance of at least $40\%$ of the vertices in the smaller part of each dissection. Furthermore, we
only use the nested dissection if separators contain at most
50 vertices. For the reduction-based
branching rules, we fall back to the default branching strategy if all
candidates have a degree of less than $\Delta - k$. In the case of twin-, funnel- and unconfined-reduction-based branching strategies we choose $k$ as $2$. For the packing-reduction-based branching rule, $k$ is set to $5$ and for the combined branching rule, $k$ is set to $4$.

\subsection{Instances}
We use instances from several sources: The ``easy'' instances used for the
PACE 2019 Challenge on Minimum Vertex Cover~\cite{dzulfikar_et_al:LIPIcs:2019:11486}. 
Complements of Maximum Clique instances from the second DIMACS Implementation Challenge~\cite{johnson1993cliques} and sparse instances from
the Stanford Network Analysis Project (SNAP)~\cite{snapnets}, the 9th DIMACS
Implementation Challenge on Shortest Paths~\cite{demetrescu2009shortest} and the
Network Data Repository~\cite{nr}. Detailed instance information can
be found in Table~\ref{table:instance}. Directed instances were converted into
undirected graphs by ignoring the direction of edges and removing duplicates.
Our original set of instances contained the first 80 PACE instances, 53 DIMACS instances and 34 sparse networks.
From these instances, we excluded all instances that (1) required no branches, (2) on which all techniques had a running time of less than $0.1$ seconds, or (3) on which no technique was able to find a solution within 10 hours.
The remaining set of instances is composed of 48 PACE instances, 37 DIMACS
instances and 16 sparse networks.

\begin{table}[htb!]	
	\scriptsize
	\caption{Number of vertices $|V|$ and edges $|E|$ for each graph.}\label{table:instance}
	\begin{center}
      \scalebox{0.75}{
		\begin{minipage}{0.29\textwidth}
			\centering
			\begin{tabular}{|l|r|r|}
				\hline
				\multicolumn{3}{|l|}{PACE \cite{dzulfikar_et_al:LIPIcs:2019:11486} instances:}                                                            \\
				\hline
				Graph                 & $|V|$              & $|E|$                                               \\
				\hline
				05 & \numprint{200} & \numprint{798}\\
				06 & \numprint{200} & \numprint{733}\\
				10 & \numprint{199} & \numprint{758}\\
				16 & \numprint{153} & \numprint{802} \\
				19 & \numprint{200} & \numprint{862}\\
				31 & \numprint{200} & \numprint{813}\\
				33 & \numprint{4410} & \numprint{6885}\\
				35 & \numprint{200} & \numprint{864}\\
				36                    & \numprint{26300}   & \numprint{41500}                                    \\
				37                    & \numprint{198}     & \numprint{808}                                      \\
				38                    & \numprint{786}     & \numprint{14024}                                    \\
				39                    & \numprint{6795}    & \numprint{10620}                                    \\
				40                    & \numprint{210}     & \numprint{625}                                      \\
				41                    & \numprint{200}     & \numprint{1023}                                     \\
				42                    & \numprint{200}     & \numprint{952}                                      \\
				43                    & \numprint{200}     & \numprint{841}                                      \\
				44                    & \numprint{200}     & \numprint{1147}                                     \\
				45                    & \numprint{200}     & \numprint{1020}                                     \\
				46                    & \numprint{200}     & \numprint{812}                                      \\
				47                    & \numprint{200}     & \numprint{1093}                                     \\
				48                    & \numprint{200}     & \numprint{1025}                                     \\
				49                    & \numprint{200}     & \numprint{933}                                      \\
				50                    & \numprint{200}     & \numprint{1025}                                     \\
				51                    & \numprint{200}     & \numprint{1098}                                     \\
				52                    & \numprint{200}     & \numprint{992}                                      \\
				53                    & \numprint{200}     & \numprint{1026}                                     \\
				54                    & \numprint{200}     & \numprint{961}                                      \\
				55                    & \numprint{200}     & \numprint{938}                                      \\
				56                    & \numprint{200}     & \numprint{1089}                                     \\
				57                    & \numprint{200}     & \numprint{1160}                                     \\
				58                    & \numprint{200}     & \numprint{1171}                                     \\
				59                    & \numprint{200}     & \numprint{961}                                      \\
				60                    & \numprint{200}     & \numprint{1118}                                     \\
				61                    & \numprint{200}     & \numprint{931}                                      \\
				62                    & \numprint{199}     & \numprint{1128}                                     \\
				63                    & \numprint{200}     & \numprint{1011}                                     \\
				64                    & \numprint{200}     & \numprint{1042}                                     \\
				65                    & \numprint{200}     & \numprint{1011}                                     \\
				66                    & \numprint{200}     & \numprint{866}                                      \\
				67                    & \numprint{200}     & \numprint{1174}                                     \\
				68                    & \numprint{200}     & \numprint{961}                                      \\
				69                    & \numprint{200}     & \numprint{1083}                                     \\
				70                    & \numprint{200}     & \numprint{860}                                      \\
				71                    & \numprint{200}     & \numprint{952}                                      \\
				72                    & \numprint{200}     & \numprint{1167}                                     \\
				73                    & \numprint{200}     & \numprint{1078}                                     \\
				74                    & \numprint{200}     & \numprint{805}                                      \\
				77                    & \numprint{200}   & \numprint{961}                                    \\
				\hline
			\end{tabular}
		\end{minipage}
		\begin{minipage}{0.7\textwidth}
			\centering
      \begin{minipage}{\textwidth}
      \centering
			\begin{tabular}{|l|r|r|}
				\hline
				\multicolumn{3}{|l|}{DIMACS \cite{johnson1993cliques} instances:}                                                          \\
				\hline
				Graph                 & $|V|$              & $|E|$                                               \\
				\hline
				C125.9 & \numprint{125} & \numprint{787} \\
				MANN\_a27 & \numprint{378} & \numprint{702} \\
				MANN\_a45 & \numprint{1035} & \numprint{1980} \\
				brock200\_1 & \numprint{200} & \numprint{5066} \\
				brock200\_2 & \numprint{200} & \numprint{10024} \\
				brock200\_3 & \numprint{200} & \numprint{7852} \\
				brock200\_4 & \numprint{200} & \numprint{6811} \\
				gen200\_p0.9\_44 & \numprint{200} & \numprint{1990} \\
				gen200\_p0.9\_55 & \numprint{200} & \numprint{1990} \\
				hamming8-4 & \numprint{256} & \numprint{11776} \\
				johnson16-2-4 & \numprint{120} & \numprint{1680} \\
				keller4 & \numprint{171} & \numprint{5100} \\
				p\_hat1000-1 & \numprint{1000} & \numprint{377247} \\
				p\_hat1000-2 & \numprint{1000} & \numprint{254701} \\
				p\_hat1500-1 & \numprint{1500} & \numprint{839327} \\
				p\_hat300-1 & \numprint{300} & \numprint{33917} \\
				p\_hat300-2 & \numprint{300} & \numprint{22922} \\
				p\_hat300-3 & \numprint{300} & \numprint{11460} \\
				p\_hat500-1 & \numprint{500} & \numprint{93181} \\
				p\_hat500-2 & \numprint{500} & \numprint{61804} \\
				p\_hat500-3 & \numprint{500} & \numprint{30950} \\
				p\_hat700-1 & \numprint{700} & \numprint{183651} \\
				p\_hat700-2 & \numprint{700} & \numprint{122922} \\
				san1000 & \numprint{1000} & \numprint{249000} \\
				san200\_0.7\_1 & \numprint{200} & \numprint{5970} \\
				san200\_0.7\_2 & \numprint{200} & \numprint{5970} \\
				san200\_0.9\_1 & \numprint{200} & \numprint{1990} \\
				san200\_0.9\_2 & \numprint{200} & \numprint{1990} \\
				san200\_0.9\_3 & \numprint{200} & \numprint{1990} \\
				san400\_0.5\_1 & \numprint{400} & \numprint{39900} \\
				san400\_0.7\_1 & \numprint{400} & \numprint{23940} \\
				san400\_0.7\_2 & \numprint{400} & \numprint{23940} \\
				san400\_0.7\_3 & \numprint{400} & \numprint{23940} \\
				sanr200\_0.7 & \numprint{200} & \numprint{6032} \\
				sanr200\_0.9 & \numprint{200} & \numprint{2037} \\
				sanr400\_0.5 & \numprint{400} & \numprint{39816} \\
				sanr400\_0.7 & \numprint{400} & \numprint{23931} \\
				\hline
			\end{tabular}
      \end{minipage}
			\vspace{1em}
			\newline
      \begin{minipage}{\textwidth}
      \centering
			\begin{tabular}{|l|r|r|c|}
				\hline			
				\multicolumn{4}{|l|}{Sparse networks:}                                                           \\
				\hline
				Graph                 & $|V|$              & $|E|$               & source                        \\			
				\hline
				as-skitter            & \numprint{1696415} & \numprint{11095298} & \cite{snapnets}               \\
				baidu-relatedpages    & \numprint{415641} & \numprint{2374044} & \cite{nr} \\
				bay                     & \numprint{321270}  & \numprint{397415}   & \cite{demetrescu2009shortest} \\
				col                   & \numprint{435666}  & \numprint{521200}   & \cite{demetrescu2009shortest} \\
				fla                   & \numprint{1070376} & \numprint{1343951}  & \cite{demetrescu2009shortest} \\
				hudong-internallink   & \numprint{1984484} & \numprint{14428382} & \cite{nr}\\
				in-2004 			  & \numprint{1382870} & \numprint{13591473} & \cite{nr}\\
				libimseti             & \numprint{220970}  & \numprint{17233144} & \cite{nr}                     \\
				musae-twitch\_DE      & \numprint{9498}    & \numprint{153138}   & \cite{snapnets}               \\
				musae-twitch\_FR      & \numprint{6549}    & \numprint{112666}   & \cite{snapnets}               \\
				petster-fs-dog & \numprint{426820} & \numprint{8543549} & \cite{nr} \\
				soc-LiveJournal1      & \numprint{4847571} & \numprint{42851237} & \cite{snapnets} \\
				web-BerkStan          & \numprint{685230}  & \numprint{6649470}  & \cite{snapnets}               \\
				web-Google            & \numprint{875713}  & \numprint{4322051}  & \cite{snapnets}               \\
				web-NotreDame         & \numprint{325730}  & \numprint{1090108}  & \cite{snapnets}               \\
				web-Stanford          & \numprint{281903}  & \numprint{1992636}  & \cite{snapnets}               \\
				\hline
			\end{tabular}
      \end{minipage}
		\end{minipage}
      }
	\end{center}
\end{table}

\FloatBarrier
\subsection{Decomposition Branching}\label{sec:experiments_decomp}



\begin{figure}[t!]
\begin{subfigure}[]{\textwidth}
	\centering
	\input{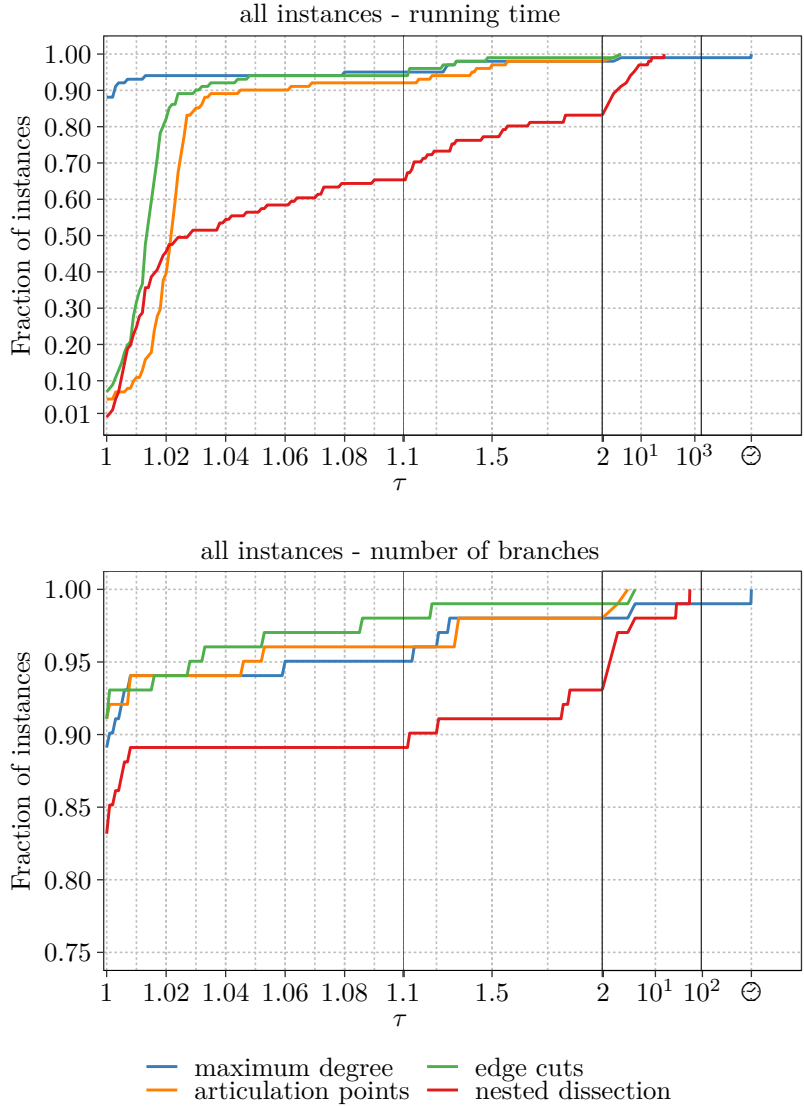}
\end{subfigure}

\caption{Performance plots for decomposition-based branching strategies}\label{fig:all_decomp}
\end{figure}

\begin{table}[t!]
  \caption{Speedup of our decomposition-based techniques over maximum degree branching.}\label{tab:summary_decomp}

  \centering
  \footnotesize
  \begin{tabular}{|l|r|r|r|r|}
    \hline
    & \multicolumn{1}{c|}{PACE} & \multicolumn{1}{c|}{DIMACS} & \multicolumn{1}{c|}{Sparse net.} & \multicolumn{1}{c|}{All Instances}                                                                                                            \\\hline
    articulation points         & \numprint{0.99}          & \numprint{0.99}             & \numprint{2.17}             & \numprint{1.20} \\
    edge cuts                   & \numprint{1.00}  & \numprint{0.99}             & \textbf{\numprint{2.29}}    & \textbf{\numprint{1.22}}  \\
    nested dissections          & \numprint{1.00}           & \numprint{0.99}          & \numprint{2.15}          & \numprint{1.21} \\
    \hline
    \end{tabular}
\end{table}
Figure~\ref{fig:all_decomp} shows performance profiles~\cite{dolan2002benchmarking} of the running time and
number of branches of
our decomposition-based branching strategies:
Let $\mathcal{T}$ be the set of all techniques we want to compare, $\mathcal{I}$
the set of instances, and $t_{T}({I})$ the running time/number of branches of
technique $T \in \mathcal{T}$ on instance $I \in \mathcal{I}$. The y-axis shows for each technique $T$ the
fraction of instances for which $t_T(I) \leq \tau \cdot \min_{T' \in
  \mathcal{T}}t_{T'}(I)$, where $\tau$ is shown on the x-axis. For $\tau = 1$,
the y-axis shows the fraction of instances on which a technique performs best.
Note that these plots compare the performance of a technique relative to the
best performing technique and do not show a ranking of all techniques.
Instances that were not finished by a technique within the time limit are marked
with \ClockLogo{}.

The running time plot in Figure~\ref{fig:all_decomp} shows that for most instances, the default strategy of branching on a
vertex of maximum degree outperforms our decomposition-based approaches.
However, for instances that have suitable candidates for decomposition, such as
sparse networks, significant speedups compared to the default strategy can be
seen. To be more specific, assigning a time of ten hours (our timeout threshold) for unfinished
instances, we achieve a total speedup\footnote{calculated by dividing the running times
to solve all instances for two algorithms, excluding instances unsolved by both
algorithms} of $\numprint{2.15}$ to $\numprint{2.29}$ over maximum degree branching for our
decomposition-based techniques on sparse networks (see
Table~\ref{tab:summary_decomp}). In particular, there is one instance (web-stanford) that causes a timeout with the
default strategy but can be solved in $8$ (articulation points) to $43$ (nested
dissections) seconds using a decomposition-based approach. Table~\ref{tab:summary_decomp} shows that overall, our
technique using edge cuts seems to be the most beneficial, achieving an overall
speedup of $22\%$ over maximum degree.
Finally, Figure~\ref{fig:all_decomp} shows that most running
times are only slightly slower than the default strategy with a few instances
showing a speedup. This is mainly because
the number of branches required to solve the instances does not change in most
cases and most of the running time
difference is caused by the overhead from
searching for branching vertices.

\subsection{Reduction Branching}
\begin{figure}[t!]
\begin{subfigure}[t!]{\textwidth}
	\centering
	\input{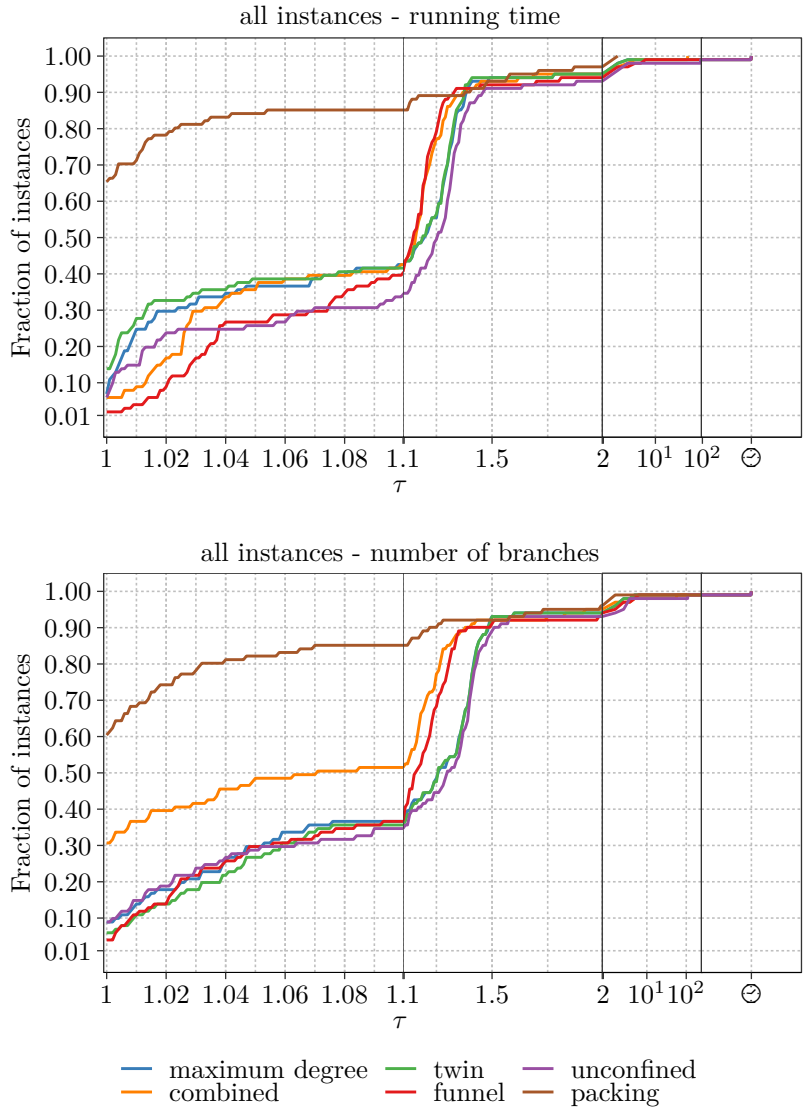}
\end{subfigure}

\caption{Performance plots for reduction-based branching strategies}\label{fig:all_reduction}
\end{figure}

Figure~\ref{fig:all_reduction} shows the performance profiles (see
Section~\ref{sec:experiments_decomp}) for our reduction-based branching strategies. Here we see that targeting the packing reduction
results in the fastest time for the most number of instances. In fact, targeting
the packing reduction performs better than maximum degree branching on all but 3
PACE instances, achieving a speedup of $34\%$ (Table~\ref{tab:summary_reduction}) on these instances. On the DIMACS
instances, performance is closer to that of maximum degree with an overall
speedup of $4\%$. On sparse networks, packing is only faster than maximum degree
branching on $6$ out of $16$ instances but still achieves an overall speedup of
$31\%$ due to being considerably faster on some of the longer running instances.
The performance of our packing-based technique might be explained by it's
property of enabling a reduction in both the including and the excluding branch,
while our other reduction-based techniques only enable a reduction in the
excluding branch.
Our funnel-based technique is faster than maximum degree branching for all but 4
of the PACE instances, resulting in a speedup of $14\%$ on these instances but
only a $2\%$ speedup over all instances due to slightly slower running times on
the other instance classes. We also show results for a strategy that targets all reduction rules
described in Section~\ref{sec:reduction_branching} (called \emph{combined}). Even though this approach
leads to the second lowest number of branches for most instances, the time required to
identify candidate vertices for all reduction rules causes too big of an
overhead to be competitive. In fact, preliminary experiments showed that the
number of branches is still small for a technique that only combines twin-,
funnel- and unconfined-based branching. Optimizing the algorithms to identify candidate
vertices could lead to making this combined strategy
competitive.



\begin{table}[t!]
  \caption{Speedup of our reduction-based techniques over maximum degree branching.}\label{tab:summary_reduction}

  \centering
  \footnotesize
  \begin{tabular}{|l|r|r|r|r|}
    \hline
    & \multicolumn{1}{c|}{PACE} & \multicolumn{1}{c|}{DIMACS} & \multicolumn{1}{c|}{Sparse net.} & \multicolumn{1}{c|}{All Instances}                                                                                                          \\
    \hline
    Twin                        & \numprint{1.00}   & \numprint{1.00}     & \numprint{0.97}  & \numprint{0.99} \\
    Funnel                      & \numprint{1.14}   & \numprint{0.99}     & \numprint{0.98}  & \numprint{1.02} \\
    Unconfined                  & \numprint{0.79}   & \numprint{1.00}     & \numprint{0.86}  &  \numprint{0.92} \\
    Packing                    & \textbf{\numprint{1.34}}  & \textbf{\numprint{1.04}}     & \textbf{\numprint{1.31}}   & \textbf{\numprint{1.16}}  \\
    Combined                    & \numprint{1.14}   & \numprint{1.03}      & \numprint{1.30}   & \numprint{1.12} \\
    \hline
    \end{tabular}
\end{table}

\section{Conclusion and Future Work}
In this work we presented several novel branching strategies for the maximum independent set problem.
Our strategies either follow a decomposition-based or reduction-rule-based approach.
The decomposition-based strategies make use of increasingly sophisticated methods of finding vertices that are likely to decompose the graph into two or more connected components.
Even though these strategies often come with a non negligible overhead, they work well for graphs that have a suitable structure, such as social networks.
For instances that still favor the default branching strategy of branching on the vertex of highest degree, our reduction-rule-based strategies provide a smaller but more consistent speedup.
These rules aim to facilitate the application of reduction rules which leads to smaller graphs that can be solved more quickly.

Overall, using one of our proposed strategies allows us to find the optimal solution the fastest for most instances tested.
However, deciding which particular strategy to use for a given instance still remains an open problem.
Finding suitable graph characteristics to do so provides an interesting opportunity for future work.
Furthermore, our experimental evaluation on a combined approach that tries to use all reduction-rule-based strategies at the same time achieves a smaller number of branches than the default strategy for a large set of instances.
However, the running time of this approach still suffers from frequent checks whether a particular vertex is a potential branching vertex.
A more sophisticated and incremental way of tracking when a vertex becomes a branching vertex might provide significant performance benefits.
In turn, this might lead to a branching strategy that consistently outperforms branching on the vertex of highest degree independent of the instance type.

\FloatBarrier


\bibliography{references}

\newpage

\appendix

\section{Detailed Experimental Results}
\label{app:detailed_results}
We now present detailed results of our experimental evaluation. Detailed tables
show running times $t$ (in seconds) and speedup $s$.
Speedups are computed by dividing the running time of maximum degree branching by the running time of the respective technique.
Timeouts are assigned a running time of ten hours. Note, that this is the same as our time limit.
We also present the aggregated speedup $s_{\text{total}}$ computed by dividing the running time of both algorithms over all instances (omitting instances were both algorithms do not finish within our time limit).
A value is highlighted in bold if it is the best one within a row.

\begin{table}[htb!]	
	\scriptsize
	\setlength{\tabcolsep}{2pt}
	\caption{Detailed results for our decomposition-based strategies on the PACE instances.}
	\begin{center}
      \scalebox{0.75}{
		\begin{tabular}{|l|r|r|r|r|}\hline
			Graph & max. deg. & \multicolumn{1}{c|}{articulation} & \multicolumn{1}{c|}{edge cuts} & \multicolumn{1}{c|}{nested dis.} \\
			\hline
			PACE & t & t (s) & t (s) & t (s) \\
			\hline
			05 & \textbf{\numprint{1.97}} & \numprint{2.00} (\numprint{0.98}) & \numprint{2.00} (\numprint{0.99}) & \numprint{2.44} (\numprint{0.81}) \\
			06 & \textbf{\numprint{0.85}} & \numprint{0.87} (\numprint{0.98}) & \numprint{0.87} (\numprint{0.98}) & \numprint{1.33} (\numprint{0.64}) \\
			10 & \textbf{\numprint{2.24}} & \numprint{2.27} (\numprint{0.99}) & \numprint{2.26} (\numprint{0.99}) & \numprint{2.66} (\numprint{0.84}) \\
			16 & \numprint{25836.77} & \numprint{26175.23} (\numprint{0.99}) & \textbf{\numprint{25763.30} (\numprint{1.00})} & \numprint{25865.40} (\numprint{1.00}) \\
			19 & \textbf{\numprint{3.17}} & \numprint{3.22} (\numprint{0.98}) & \numprint{3.18} (\numprint{0.99}) & \numprint{3.63} (\numprint{0.87}) \\
			31 & \textbf{\numprint{74.37}} & \numprint{76.03} (\numprint{0.98}) & \numprint{75.45} (\numprint{0.99}) & \numprint{74.82} (\numprint{0.99}) \\
			33 & \textbf{\numprint{1.01}} & \numprint{1.03} (\numprint{0.98}) & \numprint{1.02} (\numprint{0.99}) & \numprint{40.09} (\numprint{0.03}) \\
			35 & \textbf{\numprint{7.64}} & \numprint{7.84} (\numprint{0.97}) & \numprint{7.77} (\numprint{0.98}) & \numprint{8.13} (\numprint{0.94}) \\
			36 & \textbf{\numprint{1.84}} & \numprint{1.87} (\numprint{0.98}) & \numprint{1.85} (\numprint{0.99}) & \numprint{3.93} (\numprint{0.47}) \\
			37 & \textbf{\numprint{10.27}} & \numprint{10.48} (\numprint{0.98}) & \numprint{10.47} (\numprint{0.98}) & \numprint{10.75} (\numprint{0.96}) \\
			38 & \numprint{12.33} & \numprint{11.24} (\numprint{1.10}) & \textbf{\numprint{3.25} (\numprint{3.79})} & \numprint{15.35} (\numprint{0.80}) \\
			39 & \textbf{\numprint{93.79}} & \numprint{96.82} (\numprint{0.97}) & \numprint{95.96} (\numprint{0.98}) & \numprint{95.21} (\numprint{0.99}) \\
			40 & \textbf{\numprint{4690.64}} & \numprint{4794.37} (\numprint{0.98}) & \numprint{4758.15} (\numprint{0.99}) & \numprint{4712.57} (\numprint{1.00}) \\
			41 & \textbf{\numprint{48.56}} & \numprint{49.84} (\numprint{0.97}) & \numprint{49.39} (\numprint{0.98}) & \numprint{49.35} (\numprint{0.98}) \\
			42 & \textbf{\numprint{37.32}} & \numprint{38.11} (\numprint{0.98}) & \numprint{37.91} (\numprint{0.98}) & \numprint{37.87} (\numprint{0.99}) \\
			43 & \textbf{\numprint{175.11}} & \numprint{178.81} (\numprint{0.98}) & \numprint{177.26} (\numprint{0.99}) & \numprint{175.24} (\numprint{1.00}) \\
			44 & \textbf{\numprint{92.90}} & \numprint{95.13} (\numprint{0.98}) & \numprint{94.28} (\numprint{0.99}) & \numprint{93.40} (\numprint{0.99}) \\
			45 & \textbf{\numprint{25.41}} & \numprint{26.01} (\numprint{0.98}) & \numprint{25.73} (\numprint{0.99}) & \numprint{25.90} (\numprint{0.98}) \\
			46 & \textbf{\numprint{109.55}} & \numprint{111.95} (\numprint{0.98}) & \numprint{111.00} (\numprint{0.99}) & \numprint{110.22} (\numprint{0.99}) \\
			47 & \textbf{\numprint{58.47}} & \numprint{59.70} (\numprint{0.98}) & \numprint{59.38} (\numprint{0.98}) & \numprint{59.22} (\numprint{0.99}) \\
			48 & \textbf{\numprint{25.28}} & \numprint{25.80} (\numprint{0.98}) & \numprint{25.60} (\numprint{0.99}) & \numprint{25.80} (\numprint{0.98}) \\
			49 & \textbf{\numprint{17.80}} & \numprint{18.19} (\numprint{0.98}) & \numprint{18.10} (\numprint{0.98}) & \numprint{18.30} (\numprint{0.97}) \\
			50 & \textbf{\numprint{48.87}} & \numprint{50.01} (\numprint{0.98}) & \numprint{49.56} (\numprint{0.99}) & \numprint{49.40} (\numprint{0.99}) \\
			51 & \textbf{\numprint{56.70}} & \numprint{58.00} (\numprint{0.98}) & \numprint{57.63} (\numprint{0.98}) & \numprint{57.52} (\numprint{0.99}) \\
			52 & \textbf{\numprint{22.16}} & \numprint{22.68} (\numprint{0.98}) & \numprint{22.53} (\numprint{0.98}) & \numprint{22.69} (\numprint{0.98}) \\
			53 & \textbf{\numprint{59.88}} & \numprint{61.42} (\numprint{0.97}) & \numprint{60.77} (\numprint{0.99}) & \numprint{60.42} (\numprint{0.99}) \\
			54 & \textbf{\numprint{32.08}} & \numprint{32.89} (\numprint{0.98}) & \numprint{32.73} (\numprint{0.98}) & \numprint{32.67} (\numprint{0.98}) \\
			55 & \textbf{\numprint{6.83}} & \numprint{6.97} (\numprint{0.98}) & \numprint{6.92} (\numprint{0.99}) & \numprint{7.32} (\numprint{0.93}) \\
			56 & \textbf{\numprint{97.00}} & \numprint{99.09} (\numprint{0.98}) & \numprint{98.31} (\numprint{0.99}) & \numprint{97.80} (\numprint{0.99}) \\
			57 & \textbf{\numprint{66.01}} & \numprint{67.76} (\numprint{0.97}) & \numprint{67.18} (\numprint{0.98}) & \numprint{66.83} (\numprint{0.99}) \\
			58 & \textbf{\numprint{48.12}} & \numprint{48.83} (\numprint{0.99}) & \numprint{48.72} (\numprint{0.99}) & \numprint{48.63} (\numprint{0.99}) \\
			59 & \textbf{\numprint{13.30}} & \numprint{13.60} (\numprint{0.98}) & \numprint{13.54} (\numprint{0.98}) & \numprint{13.80} (\numprint{0.96}) \\
			60 & \textbf{\numprint{79.56}} & \numprint{81.58} (\numprint{0.98}) & \numprint{80.94} (\numprint{0.98}) & \numprint{80.23} (\numprint{0.99}) \\
			61 & \textbf{\numprint{21.91}} & \numprint{22.31} (\numprint{0.98}) & \numprint{22.26} (\numprint{0.98}) & \numprint{22.36} (\numprint{0.98}) \\
			62 & \textbf{\numprint{66.22}} & \numprint{68.48} (\numprint{0.97}) & \numprint{67.40} (\numprint{0.98}) & \numprint{66.80} (\numprint{0.99}) \\
			63 & \textbf{\numprint{69.06}} & \numprint{70.55} (\numprint{0.98}) & \numprint{69.91} (\numprint{0.99}) & \numprint{69.35} (\numprint{1.00}) \\
			64 & \textbf{\numprint{29.58}} & \numprint{30.07} (\numprint{0.98}) & \numprint{29.99} (\numprint{0.99}) & \numprint{30.09} (\numprint{0.98}) \\
			65 & \textbf{\numprint{36.84}} & \numprint{37.53} (\numprint{0.98}) & \numprint{37.28} (\numprint{0.99}) & \numprint{37.29} (\numprint{0.99}) \\
			66 & \textbf{\numprint{8.06}} & \numprint{8.28} (\numprint{0.97}) & \numprint{8.23} (\numprint{0.98}) & \numprint{8.63} (\numprint{0.93}) \\
			67 & \textbf{\numprint{122.74}} & \numprint{124.79} (\numprint{0.98}) & \numprint{124.25} (\numprint{0.99}) & \numprint{123.38} (\numprint{0.99}) \\
			68 & \textbf{\numprint{8.79}} & \numprint{8.92} (\numprint{0.99}) & \numprint{8.86} (\numprint{0.99}) & \numprint{9.24} (\numprint{0.95}) \\
			69 & \textbf{\numprint{43.11}} & \numprint{44.13} (\numprint{0.98}) & \numprint{43.85} (\numprint{0.98}) & \numprint{43.63} (\numprint{0.99}) \\
			70 & \textbf{\numprint{11.79}} & \numprint{12.00} (\numprint{0.98}) & \numprint{11.97} (\numprint{0.99}) & \numprint{12.25} (\numprint{0.96}) \\
			71 & \textbf{\numprint{36.20}} & \numprint{36.83} (\numprint{0.98}) & \numprint{36.66} (\numprint{0.99}) & \numprint{36.64} (\numprint{0.99}) \\
			72 & \textbf{\numprint{46.44}} & \numprint{47.47} (\numprint{0.98}) & \numprint{46.91} (\numprint{0.99}) & \numprint{46.86} (\numprint{0.99}) \\
			73 & \textbf{\numprint{43.02}} & \numprint{44.07} (\numprint{0.98}) & \numprint{43.77} (\numprint{0.98}) & \numprint{43.65} (\numprint{0.99}) \\
			74 & \textbf{\numprint{7.06}} & \numprint{7.24} (\numprint{0.97}) & \numprint{7.14} (\numprint{0.99}) & \numprint{7.49} (\numprint{0.94}) \\
			77 & \textbf{\numprint{13.30}} & \numprint{13.65} (\numprint{0.97}) & \numprint{13.51} (\numprint{0.98}) & \numprint{13.79} (\numprint{0.96}) \\
			\hline
			$s_{\text{total}}$ & \textbf{\numprint{1.00}} & \numprint{0.99} & \numprint{1.00} & \numprint{1.00} \\
			\hline
		\end{tabular}
        }
	\end{center}
	
\end{table}
\begin{table}
	\scriptsize
	\setlength{\tabcolsep}{2pt}
	\caption{Detailed results for our decomposition-based strategies on the DIMACS instances.}
	\begin{center}
      \scalebox{0.75}{
		\begin{tabular}{|l|r|r|r|r|}\hline
			Graph & max. deg. & \multicolumn{1}{c|}{articulation} & \multicolumn{1}{c|}{edge cuts} & \multicolumn{1}{c|}{nested dis.} \\
			\hline
			DIMACS & t & t (s) & t (s) & t (s) \\
			\hline
			C125.9 & \textbf{\numprint{0.98}} & \numprint{1.01} (\numprint{0.97}) & \numprint{1.00} (\numprint{0.98}) & \numprint{1.43} (\numprint{0.69}) \\
			MANN\_a27 & \textbf{\numprint{0.48}} & \numprint{0.49} (\numprint{0.98}) & \numprint{0.49} (\numprint{0.98}) & \numprint{0.98} (\numprint{0.49}) \\
			MANN\_a45 & \textbf{\numprint{73.80}} & \numprint{75.24} (\numprint{0.98}) & \numprint{74.93} (\numprint{0.98}) & \numprint{74.70} (\numprint{0.99}) \\
			brock200\_1 & \textbf{\numprint{137.34}} & \numprint{140.20} (\numprint{0.98}) & \numprint{137.56} (\numprint{1.00}) & \numprint{140.01} (\numprint{0.98}) \\
			brock200\_2 & \textbf{\numprint{4.59}} & \numprint{4.69} (\numprint{0.98}) & \numprint{4.70} (\numprint{0.98}) & \numprint{10.07} (\numprint{0.46}) \\
			brock200\_3 & \numprint{22.06} & \numprint{22.33} (\numprint{0.99}) & \textbf{\numprint{21.92} (\numprint{1.01})} & \numprint{26.39} (\numprint{0.84}) \\
			brock200\_4 & \textbf{\numprint{28.34}} & \numprint{28.72} (\numprint{0.99}) & \numprint{28.35} (\numprint{1.00}) & \numprint{32.48} (\numprint{0.87}) \\
			gen200\_p0.9\_44 & \textbf{\numprint{152.61}} & \numprint{156.30} (\numprint{0.98}) & \numprint{154.50} (\numprint{0.99}) & \numprint{153.49} (\numprint{0.99}) \\
			gen200\_p0.9\_55 & \textbf{\numprint{131.24}} & \numprint{134.64} (\numprint{0.97}) & \numprint{133.04} (\numprint{0.99}) & \numprint{132.58} (\numprint{0.99}) \\
			hamming8-4 & \textbf{\numprint{19.29}} & \numprint{19.65} (\numprint{0.98}) & \numprint{19.49} (\numprint{0.99}) & \numprint{25.38} (\numprint{0.76}) \\
			johnson16-2-4 & \textbf{\numprint{39.87}} & \numprint{41.17} (\numprint{0.97}) & \numprint{40.21} (\numprint{0.99}) & \numprint{40.33} (\numprint{0.99}) \\
			keller4 & \textbf{\numprint{2.62}} & \numprint{2.68} (\numprint{0.98}) & \numprint{2.65} (\numprint{0.99}) & \numprint{4.37} (\numprint{0.60}) \\
			p\_hat1000-1 & \textbf{\numprint{860.24}} & \numprint{868.71} (\numprint{0.99}) & \numprint{870.04} (\numprint{0.99}) & \numprint{906.24} (\numprint{0.95}) \\
			p\_hat1000-2 & \textbf{\numprint{33035.45}} & \numprint{33656.50} (\numprint{0.98}) & \numprint{33508.10} (\numprint{0.99}) & \numprint{33247.45} (\numprint{0.99}) \\
			p\_hat1500-1 & \textbf{\numprint{8935.77}} & \numprint{9015.15} (\numprint{0.99}) & \numprint{9015.74} (\numprint{0.99}) & \numprint{8994.28} (\numprint{0.99}) \\
			p\_hat300-1 & \textbf{\numprint{3.70}} & \numprint{3.79} (\numprint{0.98}) & \numprint{3.82} (\numprint{0.97}) & \numprint{23.94} (\numprint{0.15}) \\
			p\_hat300-2 & \textbf{\numprint{5.53}} & \numprint{5.66} (\numprint{0.98}) & \numprint{5.63} (\numprint{0.98}) & \numprint{21.76} (\numprint{0.25}) \\
			p\_hat300-3 & \numprint{189.58} & \numprint{191.06} (\numprint{0.99}) & \textbf{\numprint{188.96} (\numprint{1.00})} & \numprint{196.89} (\numprint{0.96}) \\
			p\_hat500-1 & \textbf{\numprint{38.63}} & \numprint{39.26} (\numprint{0.98}) & \numprint{39.41} (\numprint{0.98}) & \numprint{59.29} (\numprint{0.65}) \\
			p\_hat500-2 & \textbf{\numprint{96.36}} & \numprint{97.82} (\numprint{0.99}) & \numprint{97.58} (\numprint{0.99}) & \numprint{107.29} (\numprint{0.90}) \\
			p\_hat500-3 & \textbf{\numprint{14860.70}} & \numprint{14895.15} (\numprint{1.00}) & \numprint{14979.65} (\numprint{0.99}) & \numprint{14909.35} (\numprint{1.00}) \\
			p\_hat700-1 & \numprint{163.30} & \textbf{\numprint{162.84} (\numprint{1.00})} & \numprint{163.17} (\numprint{1.00}) & \numprint{177.34} (\numprint{0.92}) \\
			p\_hat700-2 & \textbf{\numprint{906.32}} & \numprint{917.87} (\numprint{0.99}) & \numprint{914.96} (\numprint{0.99}) & \numprint{917.50} (\numprint{0.99}) \\
			san1000 & \textbf{\numprint{895.34}} & \numprint{902.64} (\numprint{0.99}) & \numprint{903.38} (\numprint{0.99}) & \numprint{920.28} (\numprint{0.97}) \\
			san200\_0.7\_1 & \textbf{\numprint{10.85}} & \numprint{11.01} (\numprint{0.98}) & \numprint{10.90} (\numprint{1.00}) & \numprint{14.45} (\numprint{0.75}) \\
			san200\_0.7\_2 & \numprint{0.33} & \numprint{0.34} (\numprint{0.95}) & \textbf{\numprint{0.32} (\numprint{1.01})} & \numprint{2.34} (\numprint{0.14}) \\
			san200\_0.9\_1 & \textbf{\numprint{13.93}} & \numprint{14.37} (\numprint{0.97}) & \numprint{14.08} (\numprint{0.99}) & \numprint{14.94} (\numprint{0.93}) \\
			san200\_0.9\_2 & \textbf{\numprint{34.15}} & \numprint{34.77} (\numprint{0.98}) & \numprint{34.35} (\numprint{0.99}) & \numprint{34.90} (\numprint{0.98}) \\
			san200\_0.9\_3 & \textbf{\numprint{1069.00}} & \numprint{1094.54} (\numprint{0.98}) & \numprint{1078.09} (\numprint{0.99}) & \numprint{1071.31} (\numprint{1.00}) \\
			san400\_0.5\_1 & \textbf{\numprint{9.21}} & \numprint{9.35} (\numprint{0.98}) & \numprint{9.36} (\numprint{0.98}) & \numprint{16.76} (\numprint{0.55}) \\
			san400\_0.7\_1 & \textbf{\numprint{1125.52}} & \numprint{1139.20} (\numprint{0.99}) & \numprint{1131.38} (\numprint{0.99}) & \numprint{1130.07} (\numprint{1.00}) \\
			san400\_0.7\_2 & \numprint{3062.38} & \textbf{\numprint{3053.97} (\numprint{1.00})} & \numprint{3083.59} (\numprint{0.99}) & \numprint{3073.66} (\numprint{1.00}) \\
			san400\_0.7\_3 & \textbf{\numprint{4411.82}} & \numprint{4464.53} (\numprint{0.99}) & \numprint{4447.19} (\numprint{0.99}) & \numprint{4423.16} (\numprint{1.00}) \\
			sanr200\_0.7 & \textbf{\numprint{48.35}} & \numprint{49.51} (\numprint{0.98}) & \numprint{48.71} (\numprint{0.99}) & \numprint{52.13} (\numprint{0.93}) \\
			sanr200\_0.9 & \textbf{\numprint{679.25}} & \numprint{696.41} (\numprint{0.98}) & \numprint{688.51} (\numprint{0.99}) & \numprint{680.29} (\numprint{1.00}) \\
			sanr400\_0.5 & \textbf{\numprint{373.40}} & \numprint{374.20} (\numprint{1.00}) & \numprint{374.26} (\numprint{1.00}) & \numprint{380.08} (\numprint{0.98}) \\
			sanr400\_0.7 & \textbf{\numprint{29766.80}} & \numprint{30390.80} (\numprint{0.98}) & \numprint{30270.10} (\numprint{0.98}) & \numprint{30001.55} (\numprint{0.99}) \\
			\hline
			$s_{\text{total}}$ & \textbf{\numprint{1.00}} & \numprint{0.99} & \numprint{0.99} & \numprint{0.99} \\
			\hline
		\end{tabular}
        }
	\end{center}
	
\end{table}

\begin{table}
	\scriptsize
	\setlength{\tabcolsep}{2pt}
	\caption{Detailed results for our decomposition-based strategies on sparse networks.}
	\begin{center}
      \scalebox{0.75}{
		\begin{tabular}{|l|r|r|r|r|}\hline
			Graph & max. deg. & \multicolumn{1}{c|}{articulation} & \multicolumn{1}{c|}{edge cuts} & \multicolumn{1}{c|}{nested dis.} \\
			\hline
			Sparse net. & t & t (s) & t (s)  & t (s)  \\
			\hline
			as-skitter & \textbf{\numprint{2058.32}} & \numprint{2100.57} (\numprint{0.98}) & \numprint{2071.06} (\numprint{0.99}) & \numprint{2068.46} (\numprint{1.00}) \\
			baidu-relatedpages & \textbf{\numprint{0.82}} & \numprint{0.88} (\numprint{0.94}) & \numprint{0.86} (\numprint{0.96}) & \numprint{7.22} (\numprint{0.11}) \\
			bay & \numprint{1.68} & \numprint{1.87} (\numprint{0.90}) & \textbf{\numprint{1.31} (\numprint{1.28})} & \numprint{23.43} (\numprint{0.07}) \\
			col & \numprint{5019.93} & \numprint{4737.48} (\numprint{1.06}) & \textbf{\numprint{3872.65} (\numprint{1.30})} & \numprint{5101.46} (\numprint{0.98}) \\
			fla & \numprint{25.33} & \textbf{\numprint{23.47} (\numprint{1.08})} & \numprint{24.58} (\numprint{1.03}) & \numprint{329.42} (\numprint{0.08}) \\
			hudong-internallink & \textbf{\numprint{0.99}} & \numprint{1.55} (\numprint{0.64}) & \numprint{1.46} (\numprint{0.68}) & \numprint{1.99} (\numprint{0.50}) \\
			in-2004 & \textbf{\numprint{5.22}} & \numprint{5.46} (\numprint{0.96}) & \numprint{5.37} (\numprint{0.97}) & \numprint{16.18} (\numprint{0.32}) \\
			libimseti & \textbf{\numprint{1497.59}} & \numprint{1507.54} (\numprint{0.99}) & \numprint{1503.49} (\numprint{1.00}) & \numprint{1704.53} (\numprint{0.88}) \\
			musae-twitch\_DE & \textbf{\numprint{20906.93}} & \numprint{21470.00} (\numprint{0.97}) & \numprint{20987.30} (\numprint{1.00}) & \numprint{20949.83} (\numprint{1.00}) \\
			musae-twitch\_FR & \textbf{\numprint{37.13}} & \numprint{37.81} (\numprint{0.98}) & \numprint{37.32} (\numprint{1.00}) & \numprint{41.55} (\numprint{0.89}) \\
			petster-fs-dog & \textbf{\numprint{6.82}} & \numprint{10.21} (\numprint{0.67}) & \numprint{8.67} (\numprint{0.79}) & \numprint{12.47} (\numprint{0.55}) \\
			soc-LiveJournal1 & \textbf{\numprint{9.87}} & \numprint{11.50} (\numprint{0.86}) & \numprint{11.06} (\numprint{0.89}) & \numprint{23.91} (\numprint{0.41}) \\
			web-BerkStan & \textbf{\numprint{134.22}} & \numprint{360.88} (\numprint{0.37}) & \numprint{138.84} (\numprint{0.97}) & \numprint{207.92} (\numprint{0.65}) \\
			web-Google & \textbf{\numprint{0.61}} & \numprint{0.85} (\numprint{0.71}) & \numprint{0.68} (\numprint{0.89}) & \numprint{1.46} (\numprint{0.41}) \\
			web-NotreDame & \numprint{12.10} & \textbf{\numprint{9.07} (\numprint{1.33})} & \numprint{12.11} (\numprint{1.00}) & \numprint{48.83} (\numprint{0.25}) \\
			web-Stanford & >\numprint{36000} & \textbf{\numprint{8.38} (>\numprint{4294.84})} & \numprint{27.41} (>\numprint{1313.18}) & \numprint{42.80} (>\numprint{841.16}) \\
			\hline
			$s_{\text{total}}$ & \numprint{1.00} & \numprint{2.17} & \textbf{\numprint{2.29}} & \numprint{2.15} \\
			\hline
		\end{tabular}
        }
	\end{center}
	
\end{table}

\begin{table}
	\scriptsize
	\setlength{\tabcolsep}{2pt}
	\caption{Detailed results for our reduction-based strategies on the PACE instances.}
	\begin{center}
      \scalebox{0.75}{
		\begin{tabular}{|l|r|r|r|r|r|r|}\hline
			Graph & \multicolumn{1}{c|}{max. deg.} & \multicolumn{1}{c|}{Twin} & \multicolumn{1}{c|}{Funnel} & \multicolumn{1}{c|}{Unconfined} & \multicolumn{1}{c|}{Packing} & \multicolumn{1}{c|}{Combined}  \\
			\hline
			PACE & t & t (s) & t (s) & t (s) & t (s) & t (s) \\
			\hline
			05 & \numprint{1.97} & \numprint{1.96} (\numprint{1.01}) & \numprint{1.99} (\numprint{0.99}) & \numprint{2.04} (\numprint{0.97}) & \textbf{\numprint{1.66} (\numprint{1.19})} & \numprint{2.11} (\numprint{0.93}) \\
			06 & \numprint{0.85} & \numprint{0.85} (\numprint{1.00}) & \numprint{0.74} (\numprint{1.15}) & \numprint{0.92} (\numprint{0.92}) & \textbf{\numprint{0.67} (\numprint{1.27})} & \numprint{0.81} (\numprint{1.05}) \\
			10 & \numprint{2.24} & \numprint{2.23} (\numprint{1.01}) & \numprint{2.23} (\numprint{1.00}) & \numprint{2.32} (\numprint{0.97}) & \textbf{\numprint{1.88} (\numprint{1.19})} & \numprint{2.06} (\numprint{1.09}) \\
			16 & \numprint{25836.77} & \numprint{25856.57} (\numprint{1.00}) & \numprint{22446.13} (\numprint{1.15}) & \numprint{34642.13} (\numprint{0.75}) & \textbf{\numprint{18511.88} (\numprint{1.40})} & \numprint{22590.78} (\numprint{1.14}) \\
			19 & \numprint{3.17} & \numprint{3.14} (\numprint{1.01}) & \numprint{2.90} (\numprint{1.09}) & \numprint{3.25} (\numprint{0.98}) & \textbf{\numprint{2.60} (\numprint{1.22})} & \numprint{3.04} (\numprint{1.04}) \\
			31 & \numprint{74.37} & \numprint{74.31} (\numprint{1.00}) & \numprint{58.14} (\numprint{1.28}) & \numprint{73.23} (\numprint{1.02}) & \numprint{55.99} (\numprint{1.33}) & \textbf{\numprint{54.11} (\numprint{1.37})} \\
			33 & \numprint{1.01} & \textbf{\numprint{1.01} (\numprint{1.00})} & \numprint{1.15} (\numprint{0.88}) & \numprint{1.14} (\numprint{0.89}) & \numprint{1.02} (\numprint{0.99}) & \numprint{1.29} (\numprint{0.79}) \\
			35 & \numprint{7.64} & \numprint{7.63} (\numprint{1.00}) & \numprint{7.37} (\numprint{1.04}) & \numprint{7.90} (\numprint{0.97}) & \textbf{\numprint{6.54} (\numprint{1.17})} & \numprint{7.75} (\numprint{0.99}) \\
			36 & \textbf{\numprint{1.84}} & \numprint{1.86} (\numprint{0.99}) & \numprint{11.44} (\numprint{0.16}) & \numprint{162.22} (\numprint{0.01}) & \numprint{1.90} (\numprint{0.97}) & \numprint{75.52} (\numprint{0.02}) \\
			37 & \numprint{10.27} & \numprint{10.31} (\numprint{1.00}) & \numprint{10.27} (\numprint{1.00}) & \numprint{10.63} (\numprint{0.97}) & \textbf{\numprint{8.21} (\numprint{1.25})} & \numprint{10.90} (\numprint{0.94}) \\
			38 & \numprint{12.33} & \numprint{12.36} (\numprint{1.00}) & \numprint{11.08} (\numprint{1.11}) & \numprint{11.40} (\numprint{1.08}) & \numprint{11.44} (\numprint{1.08}) & \textbf{\numprint{10.05} (\numprint{1.23})} \\
			39 & \numprint{93.79} & \numprint{93.99} (\numprint{1.00}) & \textbf{\numprint{32.43} (\numprint{2.89})} & \numprint{127.32} (\numprint{0.74}) & \numprint{93.99} (\numprint{1.00}) & \numprint{98.25} (\numprint{0.95}) \\
			40 & \numprint{4690.64} & \numprint{4689.28} (\numprint{1.00}) & \numprint{4285.37} (\numprint{1.09}) & \numprint{4530.07} (\numprint{1.04}) & \numprint{4176.59} (\numprint{1.12}) & \textbf{\numprint{4131.79} (\numprint{1.14})} \\
			41 & \numprint{48.56} & \numprint{48.42} (\numprint{1.00}) & \numprint{42.00} (\numprint{1.16}) & \numprint{48.66} (\numprint{1.00}) & \textbf{\numprint{36.87} (\numprint{1.32})} & \numprint{38.74} (\numprint{1.25}) \\
			42 & \numprint{37.32} & \numprint{37.19} (\numprint{1.00}) & \numprint{35.69} (\numprint{1.05}) & \numprint{37.60} (\numprint{0.99}) & \textbf{\numprint{28.55} (\numprint{1.31})} & \numprint{36.07} (\numprint{1.03}) \\
			43 & \numprint{175.11} & \numprint{174.63} (\numprint{1.00}) & \numprint{158.08} (\numprint{1.11}) & \numprint{172.91} (\numprint{1.01}) & \textbf{\numprint{130.75} (\numprint{1.34})} & \numprint{154.96} (\numprint{1.13}) \\
			44 & \numprint{92.90} & \numprint{92.97} (\numprint{1.00}) & \numprint{82.64} (\numprint{1.12}) & \numprint{94.37} (\numprint{0.98}) & \textbf{\numprint{69.68} (\numprint{1.33})} & \numprint{90.20} (\numprint{1.03}) \\
			45 & \numprint{25.41} & \numprint{25.37} (\numprint{1.00}) & \numprint{25.29} (\numprint{1.01}) & \numprint{26.20} (\numprint{0.97}) & \textbf{\numprint{19.83} (\numprint{1.28})} & \numprint{26.38} (\numprint{0.96}) \\
			46 & \numprint{109.55} & \numprint{109.47} (\numprint{1.00}) & \numprint{92.61} (\numprint{1.18}) & \numprint{108.01} (\numprint{1.01}) & \textbf{\numprint{79.76} (\numprint{1.37})} & \numprint{82.72} (\numprint{1.32}) \\
			47 & \numprint{58.47} & \numprint{58.18} (\numprint{1.00}) & \numprint{53.01} (\numprint{1.10}) & \numprint{59.16} (\numprint{0.99}) & \textbf{\numprint{42.32} (\numprint{1.38})} & \numprint{52.28} (\numprint{1.12}) \\
			48 & \numprint{25.28} & \numprint{25.21} (\numprint{1.00}) & \numprint{22.65} (\numprint{1.12}) & \numprint{25.72} (\numprint{0.98}) & \textbf{\numprint{18.56} (\numprint{1.36})} & \numprint{22.93} (\numprint{1.10}) \\
			49 & \numprint{17.80} & \numprint{17.76} (\numprint{1.00}) & \numprint{16.43} (\numprint{1.08}) & \numprint{19.02} (\numprint{0.94}) & \textbf{\numprint{12.97} (\numprint{1.37})} & \numprint{16.18} (\numprint{1.10}) \\
			50 & \numprint{48.87} & \numprint{48.90} (\numprint{1.00}) & \numprint{46.07} (\numprint{1.06}) & \numprint{49.75} (\numprint{0.98}) & \textbf{\numprint{37.70} (\numprint{1.30})} & \numprint{47.09} (\numprint{1.04}) \\
			51 & \numprint{56.70} & \numprint{56.58} (\numprint{1.00}) & \numprint{51.45} (\numprint{1.10}) & \numprint{57.63} (\numprint{0.98}) & \textbf{\numprint{43.45} (\numprint{1.31})} & \numprint{50.32} (\numprint{1.13}) \\
			52 & \numprint{22.16} & \numprint{22.12} (\numprint{1.00}) & \numprint{20.56} (\numprint{1.08}) & \numprint{22.99} (\numprint{0.96}) & \textbf{\numprint{15.78} (\numprint{1.40})} & \numprint{20.82} (\numprint{1.06}) \\
			53 & \numprint{59.88} & \numprint{59.88} (\numprint{1.00}) & \numprint{54.78} (\numprint{1.09}) & \numprint{60.43} (\numprint{0.99}) & \textbf{\numprint{46.87} (\numprint{1.28})} & \numprint{55.74} (\numprint{1.07}) \\
			54 & \numprint{32.08} & \numprint{32.02} (\numprint{1.00}) & \numprint{29.29} (\numprint{1.10}) & \numprint{32.89} (\numprint{0.98}) & \textbf{\numprint{26.55} (\numprint{1.21})} & \numprint{27.76} (\numprint{1.16}) \\
			55 & \numprint{6.83} & \numprint{6.80} (\numprint{1.00}) & \numprint{6.50} (\numprint{1.05}) & \numprint{6.99} (\numprint{0.98}) & \textbf{\numprint{5.23} (\numprint{1.31})} & \numprint{6.35} (\numprint{1.08}) \\
			56 & \numprint{97.00} & \numprint{96.45} (\numprint{1.01}) & \numprint{88.78} (\numprint{1.09}) & \numprint{98.09} (\numprint{0.99}) & \textbf{\numprint{70.18} (\numprint{1.38})} & \numprint{81.46} (\numprint{1.19}) \\
			57 & \numprint{66.01} & \numprint{65.97} (\numprint{1.00}) & \numprint{57.60} (\numprint{1.15}) & \numprint{65.90} (\numprint{1.00}) & \textbf{\numprint{49.95} (\numprint{1.32})} & \numprint{52.45} (\numprint{1.26}) \\
			58 & \numprint{48.12} & \numprint{47.74} (\numprint{1.01}) & \numprint{45.82} (\numprint{1.05}) & \numprint{48.56} (\numprint{0.99}) & \textbf{\numprint{35.94} (\numprint{1.34})} & \numprint{46.62} (\numprint{1.03}) \\
			59 & \numprint{13.30} & \numprint{13.30} (\numprint{1.00}) & \numprint{12.73} (\numprint{1.04}) & \numprint{13.72} (\numprint{0.97}) & \textbf{\numprint{10.61} (\numprint{1.25})} & \numprint{12.30} (\numprint{1.08}) \\
			60 & \numprint{79.56} & \numprint{79.36} (\numprint{1.00}) & \numprint{71.73} (\numprint{1.11}) & \numprint{80.70} (\numprint{0.99}) & \textbf{\numprint{59.65} (\numprint{1.33})} & \numprint{71.85} (\numprint{1.11}) \\
			61 & \numprint{21.91} & \numprint{21.91} (\numprint{1.00}) & \numprint{20.47} (\numprint{1.07}) & \numprint{22.28} (\numprint{0.98}) & \textbf{\numprint{17.50} (\numprint{1.25})} & \numprint{21.06} (\numprint{1.04}) \\
			62 & \numprint{66.22} & \numprint{66.18} (\numprint{1.00}) & \numprint{59.16} (\numprint{1.12}) & \numprint{67.83} (\numprint{0.98}) & \textbf{\numprint{49.87} (\numprint{1.33})} & \numprint{59.64} (\numprint{1.11}) \\
			63 & \numprint{69.06} & \numprint{68.81} (\numprint{1.00}) & \numprint{61.23} (\numprint{1.13}) & \numprint{70.81} (\numprint{0.98}) & \textbf{\numprint{53.40} (\numprint{1.29})} & \numprint{58.65} (\numprint{1.18}) \\
			64 & \numprint{29.58} & \numprint{29.38} (\numprint{1.01}) & \numprint{26.96} (\numprint{1.10}) & \numprint{29.46} (\numprint{1.00}) & \textbf{\numprint{22.35} (\numprint{1.32})} & \numprint{26.78} (\numprint{1.10}) \\
			65 & \numprint{36.84} & \numprint{36.72} (\numprint{1.00}) & \numprint{33.42} (\numprint{1.10}) & \numprint{37.93} (\numprint{0.97}) & \textbf{\numprint{28.23} (\numprint{1.30})} & \numprint{31.17} (\numprint{1.18}) \\
			66 & \numprint{8.06} & \numprint{8.06} (\numprint{1.00}) & \numprint{7.47} (\numprint{1.08}) & \numprint{8.21} (\numprint{0.98}) & \textbf{\numprint{6.21} (\numprint{1.30})} & \numprint{7.97} (\numprint{1.01}) \\
			67 & \numprint{122.74} & \numprint{122.34} (\numprint{1.00}) & \numprint{113.33} (\numprint{1.08}) & \numprint{123.58} (\numprint{0.99}) & \textbf{\numprint{95.55} (\numprint{1.28})} & \numprint{112.43} (\numprint{1.09}) \\
			68 & \numprint{8.79} & \numprint{8.75} (\numprint{1.00}) & \numprint{8.92} (\numprint{0.99}) & \numprint{8.94} (\numprint{0.98}) & \textbf{\numprint{6.69} (\numprint{1.31})} & \numprint{8.57} (\numprint{1.03}) \\
			69 & \numprint{43.11} & \numprint{43.11} (\numprint{1.00}) & \numprint{38.46} (\numprint{1.12}) & \numprint{44.18} (\numprint{0.98}) & \textbf{\numprint{33.88} (\numprint{1.27})} & \numprint{39.86} (\numprint{1.08}) \\
			70 & \numprint{11.79} & \numprint{11.73} (\numprint{1.00}) & \numprint{10.09} (\numprint{1.17}) & \numprint{12.22} (\numprint{0.96}) & \textbf{\numprint{9.71} (\numprint{1.21})} & \numprint{9.76} (\numprint{1.21}) \\
			71 & \numprint{36.20} & \numprint{35.91} (\numprint{1.01}) & \numprint{32.22} (\numprint{1.12}) & \numprint{35.37} (\numprint{1.02}) & \textbf{\numprint{27.23} (\numprint{1.33})} & \numprint{33.39} (\numprint{1.08}) \\
			72 & \numprint{46.44} & \numprint{46.18} (\numprint{1.01}) & \numprint{41.66} (\numprint{1.11}) & \numprint{46.68} (\numprint{0.99}) & \textbf{\numprint{36.28} (\numprint{1.28})} & \numprint{41.86} (\numprint{1.11}) \\
			73 & \numprint{43.02} & \numprint{43.00} (\numprint{1.00}) & \numprint{40.38} (\numprint{1.07}) & \numprint{43.77} (\numprint{0.98}) & \textbf{\numprint{31.91} (\numprint{1.35})} & \numprint{43.51} (\numprint{0.99}) \\
			74 & \numprint{7.06} & \numprint{7.06} (\numprint{1.00}) & \numprint{6.67} (\numprint{1.06}) & \numprint{7.86} (\numprint{0.90}) & \textbf{\numprint{5.48} (\numprint{1.29})} & \numprint{6.96} (\numprint{1.01}) \\
			77 & \numprint{13.30} & \numprint{13.25} (\numprint{1.00}) & \numprint{12.74} (\numprint{1.04}) & \numprint{13.80} (\numprint{0.96}) & \textbf{\numprint{10.61} (\numprint{1.25})} & \numprint{12.31} (\numprint{1.08}) \\
			\hline
			$s_{\text{total}}$ & \numprint{1.00} & \numprint{1.00} & \numprint{1.14} & \numprint{0.79} & \textbf{\numprint{1.34}} & \numprint{1.14} \\
			\hline
		\end{tabular}
        }
	\end{center}
	
\end{table}

\begin{table}
	\scriptsize
	\setlength{\tabcolsep}{2pt}
	\caption{Detailed results for our reduction-based strategies on the DIMACS instances.}
	\begin{center}
      \scalebox{0.75}{
		\begin{tabular}{|l|r|r|r|r|r|r|}\hline
			Graph & \multicolumn{1}{c|}{max. deg.} & \multicolumn{1}{c|}{Twin} & \multicolumn{1}{c|}{Funnel} & \multicolumn{1}{c|}{Unconfined} & \multicolumn{1}{c|}{Packing} & \multicolumn{1}{c|}{Combined}  \\
			\hline
			DIMACS & t & t (s) & t (s) & t (s) & t (s) & t (s) \\
			\hline
			C125.9 & \numprint{0.98} & \numprint{0.98} (\numprint{1.00}) & \numprint{0.92} (\numprint{1.07}) & \numprint{0.98} (\numprint{1.00}) & \textbf{\numprint{0.85} (\numprint{1.15})} & \numprint{0.91} (\numprint{1.08}) \\
			MANN\_a27 & \numprint{0.48} & \numprint{0.48} (\numprint{1.00}) & \numprint{0.57} (\numprint{0.85}) & \numprint{0.52} (\numprint{0.92}) & \textbf{\numprint{0.48} (\numprint{1.01})} & \numprint{0.59} (\numprint{0.82}) \\
			MANN\_a45 & \numprint{73.80} & \numprint{73.76} (\numprint{1.00}) & \numprint{83.81} (\numprint{0.88}) & \numprint{78.58} (\numprint{0.94}) & \textbf{\numprint{71.86} (\numprint{1.03})} & \numprint{85.47} (\numprint{0.86}) \\
			brock200\_1 & \numprint{137.34} & \numprint{136.98} (\numprint{1.00}) & \numprint{140.15} (\numprint{0.98}) & \numprint{137.32} (\numprint{1.00}) & \textbf{\numprint{135.14} (\numprint{1.02})} & \numprint{138.64} (\numprint{0.99}) \\
			brock200\_2 & \numprint{4.59} & \numprint{4.60} (\numprint{1.00}) & \numprint{4.71} (\numprint{0.98}) & \numprint{4.59} (\numprint{1.00}) & \textbf{\numprint{4.58} (\numprint{1.00})} & \numprint{4.70} (\numprint{0.98}) \\
			brock200\_3 & \numprint{22.06} & \numprint{21.78} (\numprint{1.01}) & \numprint{22.38} (\numprint{0.99}) & \numprint{21.85} (\numprint{1.01}) & \textbf{\numprint{21.76} (\numprint{1.01})} & \numprint{22.46} (\numprint{0.98}) \\
			brock200\_4 & \numprint{28.34} & \textbf{\numprint{28.15} (\numprint{1.01})} & \numprint{29.09} (\numprint{0.97}) & \numprint{28.16} (\numprint{1.01}) & \numprint{28.25} (\numprint{1.00}) & \numprint{29.24} (\numprint{0.97}) \\
			gen200\_p0.9\_44 & \numprint{152.61} & \numprint{152.40} (\numprint{1.00}) & \numprint{136.94} (\numprint{1.11}) & \numprint{169.47} (\numprint{0.90}) & \textbf{\numprint{132.81} (\numprint{1.15})} & \numprint{149.63} (\numprint{1.02}) \\
			gen200\_p0.9\_55 & \numprint{131.24} & \numprint{131.20} (\numprint{1.00}) & \numprint{125.61} (\numprint{1.04}) & \numprint{127.51} (\numprint{1.03}) & \numprint{102.10} (\numprint{1.29}) & \textbf{\numprint{50.64} (\numprint{2.59})} \\
			hamming8-4 & \numprint{19.29} & \numprint{19.30} (\numprint{1.00}) & \numprint{19.78} (\numprint{0.98}) & \textbf{\numprint{19.12} (\numprint{1.01})} & \numprint{19.35} (\numprint{1.00}) & \numprint{19.67} (\numprint{0.98}) \\
			johnson16-2-4 & \numprint{39.87} & \numprint{39.79} (\numprint{1.00}) & \numprint{41.63} (\numprint{0.96}) & \numprint{41.40} (\numprint{0.96}) & \textbf{\numprint{38.70} (\numprint{1.03})} & \numprint{43.09} (\numprint{0.93}) \\
			keller4 & \numprint{2.62} & \numprint{2.62} (\numprint{1.00}) & \numprint{2.68} (\numprint{0.98}) & \numprint{2.63} (\numprint{1.00}) & \textbf{\numprint{2.58} (\numprint{1.02})} & \numprint{2.65} (\numprint{0.99}) \\
			p\_hat1000-1 & \numprint{860.24} & \textbf{\numprint{859.74} (\numprint{1.00})} & \numprint{870.92} (\numprint{0.99}) & \numprint{873.91} (\numprint{0.98}) & \numprint{862.77} (\numprint{1.00}) & \numprint{871.60} (\numprint{0.99}) \\
			p\_hat1000-2 & \numprint{33035.45} & \numprint{33314.15} (\numprint{0.99}) & \numprint{32999.15} (\numprint{1.00}) & \numprint{32812.80} (\numprint{1.01}) & \textbf{\numprint{30913.22} (\numprint{1.07})} & \numprint{31202.52} (\numprint{1.06}) \\
			p\_hat1500-1 & \numprint{8935.77} & \textbf{\numprint{8935.50} (\numprint{1.00})} & \numprint{9009.69} (\numprint{0.99}) & \numprint{8954.18} (\numprint{1.00}) & \numprint{8958.19} (\numprint{1.00}) & \numprint{9046.97} (\numprint{0.99}) \\
			p\_hat300-1 & \numprint{3.70} & \numprint{3.69} (\numprint{1.00}) & \numprint{3.78} (\numprint{0.98}) & \numprint{3.69} (\numprint{1.00}) & \textbf{\numprint{3.68} (\numprint{1.00})} & \numprint{3.78} (\numprint{0.98}) \\
			p\_hat300-2 & \numprint{5.53} & \numprint{5.53} (\numprint{1.00}) & \numprint{5.68} (\numprint{0.97}) & \numprint{5.54} (\numprint{1.00}) & \textbf{\numprint{5.48} (\numprint{1.01})} & \numprint{5.63} (\numprint{0.98}) \\
			p\_hat300-3 & \numprint{189.58} & \numprint{187.77} (\numprint{1.01}) & \numprint{189.16} (\numprint{1.00}) & \numprint{185.68} (\numprint{1.02}) & \textbf{\numprint{175.01} (\numprint{1.08})} & \numprint{179.53} (\numprint{1.06}) \\
			p\_hat500-1 & \numprint{38.63} & \numprint{38.70} (\numprint{1.00}) & \numprint{39.36} (\numprint{0.98}) & \numprint{39.03} (\numprint{0.99}) & \textbf{\numprint{38.61} (\numprint{1.00})} & \numprint{39.34} (\numprint{0.98}) \\
			p\_hat500-2 & \numprint{96.36} & \numprint{96.39} (\numprint{1.00}) & \numprint{97.87} (\numprint{0.98}) & \numprint{96.21} (\numprint{1.00}) & \textbf{\numprint{95.08} (\numprint{1.01})} & \numprint{96.96} (\numprint{0.99}) \\
			p\_hat500-3 & \numprint{14860.70} & \numprint{14887.15} (\numprint{1.00}) & \numprint{14624.90} (\numprint{1.02}) & \numprint{14765.90} (\numprint{1.01}) & \textbf{\numprint{13429.92} (\numprint{1.11})} & \numprint{13712.38} (\numprint{1.08}) \\
			p\_hat700-1 & \numprint{163.30} & \textbf{\numprint{160.75} (\numprint{1.02})} & \numprint{163.63} (\numprint{1.00}) & \numprint{160.81} (\numprint{1.02}) & \numprint{163.24} (\numprint{1.00}) & \numprint{163.31} (\numprint{1.00}) \\
			p\_hat700-2 & \numprint{906.32} & \numprint{908.46} (\numprint{1.00}) & \numprint{914.56} (\numprint{0.99}) & \numprint{906.78} (\numprint{1.00}) & \textbf{\numprint{866.08} (\numprint{1.05})} & \numprint{879.99} (\numprint{1.03}) \\
			san1000 & \textbf{\numprint{895.34}} & \numprint{898.16} (\numprint{1.00}) & \numprint{906.21} (\numprint{0.99}) & \numprint{901.40} (\numprint{0.99}) & \numprint{913.29} (\numprint{0.98}) & \numprint{932.29} (\numprint{0.96}) \\
			san200\_0.7\_1 & \numprint{10.85} & \textbf{\numprint{10.78} (\numprint{1.01})} & \numprint{11.01} (\numprint{0.99}) & \numprint{10.91} (\numprint{0.99}) & \numprint{10.93} (\numprint{0.99}) & \numprint{11.06} (\numprint{0.98}) \\
			san200\_0.7\_2 & \numprint{0.33} & \numprint{0.32} (\numprint{1.04}) & \numprint{0.33} (\numprint{0.98}) & \textbf{\numprint{0.31} (\numprint{1.07})} & \numprint{0.32} (\numprint{1.01}) & \numprint{0.33} (\numprint{0.99}) \\
			san200\_0.9\_1 & \numprint{13.93} & \numprint{13.90} (\numprint{1.00}) & \numprint{13.35} (\numprint{1.04}) & \textbf{\numprint{4.94} (\numprint{2.82})} & \numprint{12.03} (\numprint{1.16}) & \numprint{12.13} (\numprint{1.15}) \\
			san200\_0.9\_2 & \numprint{34.15} & \numprint{33.87} (\numprint{1.01}) & \numprint{21.46} (\numprint{1.59}) & \numprint{12.32} (\numprint{2.77}) & \numprint{15.80} (\numprint{2.16}) & \textbf{\numprint{10.01} (\numprint{3.41})} \\
			san200\_0.9\_3 & \numprint{1069.00} & \numprint{1068.17} (\numprint{1.00}) & \numprint{1016.33} (\numprint{1.05}) & \numprint{639.01} (\numprint{1.67}) & \numprint{843.40} (\numprint{1.27}) & \textbf{\numprint{600.71} (\numprint{1.78})} \\
			san400\_0.5\_1 & \numprint{9.21} & \numprint{9.21} (\numprint{1.00}) & \numprint{9.37} (\numprint{0.98}) & \textbf{\numprint{9.13} (\numprint{1.01})} & \numprint{9.24} (\numprint{1.00}) & \numprint{9.37} (\numprint{0.98}) \\
			san400\_0.7\_1 & \numprint{1125.52} & \textbf{\numprint{1121.99} (\numprint{1.00})} & \numprint{1146.32} (\numprint{0.98}) & \numprint{1125.12} (\numprint{1.00}) & \numprint{1132.10} (\numprint{0.99}) & \numprint{1151.14} (\numprint{0.98}) \\
			san400\_0.7\_2 & \numprint{3062.38} & \numprint{3063.23} (\numprint{1.00}) & \numprint{3066.62} (\numprint{1.00}) & \numprint{3463.29} (\numprint{0.88}) & \textbf{\numprint{3048.94} (\numprint{1.00})} & \numprint{3489.72} (\numprint{0.88}) \\
			san400\_0.7\_3 & \numprint{4411.82} & \numprint{4405.26} (\numprint{1.00}) & \numprint{4487.18} (\numprint{0.98}) & \textbf{\numprint{4398.18} (\numprint{1.00})} & \numprint{4497.81} (\numprint{0.98}) & \numprint{4521.80} (\numprint{0.98}) \\
			sanr200\_0.7 & \numprint{48.35} & \textbf{\numprint{48.34} (\numprint{1.00})} & \numprint{50.09} (\numprint{0.97}) & \numprint{48.41} (\numprint{1.00}) & \numprint{48.49} (\numprint{1.00}) & \numprint{50.25} (\numprint{0.96}) \\
			sanr200\_0.9 & \numprint{679.25} & \numprint{679.65} (\numprint{1.00}) & \numprint{633.59} (\numprint{1.07}) & \numprint{664.95} (\numprint{1.02}) & \textbf{\numprint{531.48} (\numprint{1.28})} & \numprint{567.49} (\numprint{1.20}) \\
			sanr400\_0.5 & \numprint{373.40} & \textbf{\numprint{370.59} (\numprint{1.01})} & \numprint{376.93} (\numprint{0.99}) & \numprint{377.71} (\numprint{0.99}) & \numprint{370.72} (\numprint{1.01}) & \numprint{376.10} (\numprint{0.99}) \\
			sanr400\_0.7 & \numprint{29766.80} & \numprint{29838.40} (\numprint{1.00}) & \numprint{30466.35} (\numprint{0.98}) & \numprint{29844.65} (\numprint{1.00}) & \textbf{\numprint{29473.60} (\numprint{1.01})} & \numprint{30242.80} (\numprint{0.98}) \\
			\hline
			$s_{\text{total}}$ & \numprint{1.00} & \numprint{1.00} & \numprint{0.99} & \numprint{1.00} & \textbf{\numprint{1.04}} & \numprint{1.03} \\
			\hline
		\end{tabular}
        }
	\end{center}
\end{table}

\begin{table}
	\scriptsize
	\setlength{\tabcolsep}{2pt}
	\caption{Detailed results for our reduction-based strategies on sparse networks.}
	\begin{center}
      \scalebox{0.75}{
		\begin{tabular}{|l|r|r|r|r|r|r|}\hline
			Graph & \multicolumn{1}{c|}{max. deg.} & \multicolumn{1}{c|}{Twin} & \multicolumn{1}{c|}{Funnel} & \multicolumn{1}{c|}{Unconfined} & \multicolumn{1}{c|}{Packing} & \multicolumn{1}{c|}{Combined}  \\
			\hline
			Sparse net. & t & t (s) & t (s) & t (s) & t (s) & t (s) \\
			\hline
			as-skitter & \numprint{2058.32} & \numprint{2054.41} (\numprint{1.00}) & \numprint{1849.79} (\numprint{1.11}) & \numprint{1977.94} (\numprint{1.04}) & \textbf{\numprint{1681.87} (\numprint{1.22})} & \numprint{1704.73} (\numprint{1.21}) \\
			baidu-relatedpages & \numprint{0.82} & \textbf{\numprint{0.80} (\numprint{1.02})} & \numprint{0.84} (\numprint{0.97}) & \numprint{0.85} (\numprint{0.97}) & \numprint{0.83} (\numprint{0.98}) & \numprint{0.93} (\numprint{0.88}) \\
			bay & \textbf{\numprint{1.68}} & \numprint{1.68} (\numprint{1.00}) & \numprint{8.22} (\numprint{0.20}) & \numprint{4.71} (\numprint{0.36}) & \numprint{1.89} (\numprint{0.89}) & \numprint{8.38} (\numprint{0.20}) \\
			col & \textbf{\numprint{5019.93}} & \numprint{5752.08} (\numprint{0.87}) & \numprint{5416.72} (\numprint{0.93}) & \numprint{8187.80} (\numprint{0.61}) & \numprint{9370.05} (\numprint{0.54}) & \numprint{5924.10} (\numprint{0.85}) \\
			fla & \textbf{\numprint{25.33}} & \numprint{25.41} (\numprint{1.00}) & \numprint{45.62} (\numprint{0.56}) & \numprint{76.60} (\numprint{0.33}) & \numprint{34.78} (\numprint{0.73}) & \numprint{42.75} (\numprint{0.59}) \\
			hudong-internallink & \textbf{\numprint{0.99}} & \numprint{1.31} (\numprint{0.76}) & \numprint{1.27} (\numprint{0.78}) & \numprint{1.21} (\numprint{0.82}) & \numprint{1.55} (\numprint{0.64}) & \numprint{1.12} (\numprint{0.88}) \\
			in-2004 & \numprint{5.22} & \textbf{\numprint{4.88} (\numprint{1.07})} & \numprint{5.25} (\numprint{0.99}) & \numprint{10.85} (\numprint{0.48}) & \numprint{5.50} (\numprint{0.95}) & \numprint{10.73} (\numprint{0.49}) \\
			libimseti & \numprint{1497.59} & \numprint{1452.17} (\numprint{1.03}) & \numprint{1620.09} (\numprint{0.92}) & \textbf{\numprint{1440.71} (\numprint{1.04})} & \numprint{1476.25} (\numprint{1.01}) & \numprint{1706.07} (\numprint{0.88}) \\
			musae-twitch\_DE & \numprint{20906.93} & \numprint{20996.87} (\numprint{1.00}) & \numprint{21190.67} (\numprint{0.99}) & \numprint{22650.53} (\numprint{0.92}) & \textbf{\numprint{19345.03} (\numprint{1.08})} & \numprint{23006.50} (\numprint{0.91}) \\
			musae-twitch\_FR & \numprint{37.13} & \numprint{37.04} (\numprint{1.00}) & \numprint{38.58} (\numprint{0.96}) & \numprint{41.15} (\numprint{0.90}) & \textbf{\numprint{35.60} (\numprint{1.04})} & \numprint{42.46} (\numprint{0.87}) \\
			petster-fs-dog & \numprint{6.82} & \textbf{\numprint{6.62} (\numprint{1.03})} & \numprint{8.16} (\numprint{0.84}) & \numprint{8.66} (\numprint{0.79}) & \numprint{9.68} (\numprint{0.70}) & \numprint{9.20} (\numprint{0.74}) \\
			soc-LiveJournal1 & \numprint{9.87} & \textbf{\numprint{6.64} (\numprint{1.49})} & \numprint{9.57} (\numprint{1.03}) & \numprint{9.49} (\numprint{1.04}) & \numprint{11.33} (\numprint{0.87}) & \numprint{10.69} (\numprint{0.92}) \\
			web-BerkStan & \numprint{134.22} & \numprint{135.47} (\numprint{0.99}) & \textbf{\numprint{122.30} (\numprint{1.10})} & \numprint{146.94} (\numprint{0.91}) & \numprint{123.60} (\numprint{1.09}) & \numprint{174.07} (\numprint{0.77}) \\
			web-Google & \numprint{0.61} & \textbf{\numprint{0.53} (\numprint{1.15})} & \numprint{0.69} (\numprint{0.87}) & \numprint{0.68} (\numprint{0.89}) & \numprint{0.78} (\numprint{0.78}) & \numprint{0.68} (\numprint{0.89}) \\
			web-NotreDame & \textbf{\numprint{12.10}} & \numprint{12.63} (\numprint{0.96}) & \numprint{15.23} (\numprint{0.79}) & \numprint{12.38} (\numprint{0.98}) & \numprint{14.09} (\numprint{0.86}) & \numprint{17.52} (\numprint{0.69}) \\
			web-Stanford & >\numprint{36000} & >\numprint{36000} & >\numprint{36000} & >\numprint{36000} & \textbf{\numprint{17886.35} (>\numprint{2.01})} & \numprint{17989.97} (>\numprint{2.00}) \\
			\hline
			$s_{\text{total}}$ & \numprint{1.00} & \numprint{0.97} & \numprint{0.98} & \numprint{0.86} & \textbf{\numprint{1.31}} & \numprint{1.30} \\
			\hline
		\end{tabular}
        }
	\end{center}
	
\end{table}
\end{document}